\documentclass[fleqn, usenatbib]{mnras}

\usepackage{amsmath}
\usepackage{amsfonts}
\usepackage{amsbsy}
\usepackage{amssymb}
\usepackage{makecell}
\usepackage[T1]{fontenc}
\usepackage[utf8]{inputenc}
\usepackage{ae,aecompl}
\usepackage{subfigure}
\usepackage{verbatim}
\usepackage{graphicx}
\usepackage{xcolor}
\usepackage{multirow}
\usepackage{url}
\usepackage{lastpage}

\usepackage{acronym} 
\usepackage{soul} 
\usepackage{mathtools} 
\usepackage{cleveref} 

\newacro{AGN}{active galactic nucleus}
\newcommand{\AGN}{\ac{AGN}}
\newacroplural{AGN}{active galactic nuclei}
\newcommand{\AGNs}{\acp{AGN}}

\newacro{ISM}{interstellar medium}
\newcommand{\ISM}{\ac{ISM}}

\newacro{YSO}{young stellar object}

\newacroplural{YSO}{young stellar objects}

\newacro{SI}{streaming instability}
\newcommand{\SI}{\ac{SI}}

\newacro{RDI}{resonant drag instability}
\newcommand{\RDI}{\ac{RDI}}
\newacroplural{RDI}{resonant drag instabilities}
\newcommand{\RDIs}{\acp{RDI}}

\newacro{PPD}{proto-planetary disc}

\newacro{ODE}{ordinary differential equation}
\newcommand{\ODE}{\ac{ODE}}
\newacroplural{ODE}{ordinary differential equations}

\newacro{IVP}{initial value problem}
\newcommand{\IVP}{\ac{IVP}}

\newcommand\bb[1]{\mbox{\boldmath{$#1$}}}

\newcommand{\pD}[2]{\partial_{#1} #2} 
\newcommand{\D}[2]{{\rm d}_{#1} #2} 
\newcommand\grad{\bb{\nabla}} 
\newcommand{\DD}[2]{\frac{{\rm d}^2 #2}{{\rm d} #1^2}}

\newcommand\bcdot{\,\bb{\cdot}\,}

\newcommand{\dtg}{\mu} 
\newcommand{\cs}{c_{\mathrm{s}}} 

\newcommand{\bx}{\mathbf{x}} 
\newcommand{\bk}{\mathbf{k}} 

\newcommand{\ug}{u_{g}} 
\newcommand{\bug}{\ensuremath{\bb{u}_{g}}} 
\newcommand{\rhog}{\rho_{g}} 
\newcommand{\bud}{\ensuremath{\bb{u}_{d}}} 
\newcommand{\rhod}{\rho_{d}} 

\newcommand{\dtgo}{\overline{\dtg}} 
\newcommand{\bugo}{\ensuremath{\overline{\bb{u}}}_{g}} 
\newcommand{\rhogo}{\overline{\rho_{g}}} 
\newcommand{\budo}{\ensuremath{\overline{\bb{u}}}_{d}} 
\newcommand{\rhodo}{\overline{\rho_{d}}} 
\newcommand{\bv}{\ensuremath{\bb{v}}} 

\newcommand{\bugi}{\ensuremath{\bb{u}_{g}'}} 
\newcommand{\rgi}{\varrho_{g}'} 
\newcommand{\budi}{\ensuremath{\bb{u}_{d}'}} 
\newcommand{\rdi}{\varrho_{d}'} 

\newcommand{\hrg}{\widehat{\varrho}_{g}} 
\newcommand{\hrd}{\widehat{\varrho}_{d}} 

\newcommand{\trg}{\widetilde{\varrho}_{g}} 
\newcommand{\trd}{\widetilde{\varrho}_{d}} 

\newcommand{\hug}{\ensuremath{\widehat{u}_{g}}} 
\newcommand{\tud}{\ensuremath{\widetilde{u}_{d}}} 
\newcommand{\tug}{\ensuremath{\widetilde{u}_{g}}} 

\newcommand{\hud}{\ensuremath{\widehat{u}_{d}}} 

\newcommand{\omegaO}{\omega_{0}} 
\newcommand{\omegaI}{\omega_{1}} 
\newcommand{\hugO}{\widehat{u}_{g, 1}} 
\newcommand{\hugI}{\widehat{u}_{g, 2}} 
\newcommand{\hrgO}{\widehat{\varrho}_{g, 1}} 
\newcommand{\hrgI}{\widehat{\varrho}_{g, 2}} 
\newcommand{\hudO}{\widehat{u}_{d, 1}} 
\newcommand{\hrdO}{\widehat{\varrho}_{d, 0}} 

\newcommand{\vpara}{v_{\parallel}} 
\newcommand{\bvperp}{\mathbf{v}_{\perp}} 
\newcommand{\vperp}{v_{\perp}} 

\newcommand{\tugpara}{\widetilde{u}_{g, \parallel}} 
\newcommand{\tbugperp}{\bb{\widetilde{u}}_{g, \perp}} 
\newcommand{\tudpara}{\widetilde{u}_{d, \parallel}} 
\newcommand{\tbudperp}{\bb{\widetilde{u}}_{d, \perp}} 

\newcommand{\hbugperp}{\bb{\widehat{u}}_{g, \perp}} 
\newcommand{\hbudperp}{\bb{\widehat{u}}_{d, \perp}} 

\newcommand{\hugparaO}{\widehat{u}_{g, \parallel, 1}} 
\newcommand{\hudparaO}{\widehat{u}_{d, \parallel, 1}} 
\newcommand{\hbudperpO}{\bb{\widehat{u}}_{d, \perp, 1}} 

\newcommand{\Omegap}{\Omega'} 
\newcommand{\tOmega}{\tilde{\Omega}} 
\newcommand{\tOmegap}{\tilde{\Omega}'} 

\newcommand{\e}{\mathrm{e}} 

\title[Explaining the acoustic RDI]{A physical picture for the acoustic resonant drag instability}
\author[N.~Magnan, T.~Heinemann \& H.~Latter]{
    Nathan Magnan$^{1}$\thanks{Contact e-mail: \href{mailto:nmtm2@cam.ac.uk}{nathan.magnan@maths.cam.ac.uk}}, 
    Tobias Heinemann$^{2}$, 
    and Henrik N.~Latter$^{1}$ 
    \\
    $^{1}$DAMTP, University of Cambridge, CMS, Wilberforce Road, Cambridge CB3 0WA, UK. \\
    $^{2}$Niels Bohr International Academy, Niels Bohr Institute, Blegdamsvej 17, 2100 Copenhagen, Denmark
    }

\begin{document}

\label{firstpage}
\pagerange{\pageref{firstpage}--\pageref{lastpage}}

\maketitle

\begin{abstract}
\noindent Mixtures of gas and dust are pervasive in the universe, from AGN and molecular clouds to proto-planetary discs. When the two species drift relative to each other, a large class of instabilities can arise, called 'resonant drag instabilities' (RDIs). The most famous RDI is the streaming instability, which plays an important role in planet formation. On the other hand, acoustic RDIs, the simplest kind, feature in the winds of cool stars, AGN, or starburst regions. Unfortunately, owing to the complicated dynamics of two coupled fluids (gas and dust), the underlying physics of most RDIs is mysterious. 
In this paper, we develop a clear physical picture of how the acoustic RDI arises and support this explanation with transparent mathematics. We find that the acoustic RDI is built on two coupled mechanisms. In the first, the converging flows of a sound wave concentrate dust. In the second, a drifting dust clump excites sound waves. These processes feed into each other at resonance, thereby closing an unstable feedback loop. This physical picture helps decide where and when RDIs are most likely to happen, and what can suppress them. Additionally, we find that the acoustic RDI remains strong far from resonance. This second result suggests that one can simulate RDIs without having to fine-tune the dimensions of the numerical domain. 
\end{abstract}

\begin{keywords}
    hydrodynamics  --- instabilities --- turbulence --- waves
\end{keywords}

\defcitealias{SquireHopkins18b}{SH18b}
\defcitealias{HopkinsSquire18a}{HS18}
\defcitealias{SquireHopkins20}{SH20}

\section{Introduction}
\label{sec:intro}

Mixtures of gas and dust are omnipresent in the universe. They appear in the winds of \AGNs\ \citep{HickoxAlexander18}, molecular clouds in the \ISM\ \citep{vanDishoeck04}, the proto-planetary and debris discs surrounding young stars \citep{Andrews20, Hughes+18}, and the atmospheres of AGB stars \citep{Decin21}. Aerosols are also intensively studied by other communities \citep{Chan+22}; for instance, atmospheric physicists are interested in applications to droplet formation in clouds, and engineers to turbulent combustion \citep{ToschiBodenschatz09}.

When the body and surface forces applied to the two species differ, the dust can drift relative to the gas. In astrophysical contexts, this may be due to preferential dust acceleration by radiation pressure (\AGN\ winds, cool stars, molecular clouds), or by pressure gradients working solely on the gas (proto-planetary discs). In any system where there is such a drift, the dust can play an important dynamical role by triggering certain hydrodynamical instabilities. These instabilities were first studied by \cite{Morris93} and \cite{Deguchi97} in the context of cool-star winds.\footnote{see \cite{HopkinsSquire18a} -- hereafter \citetalias{HopkinsSquire18a} -- for more details.} But the most famous is the \SI\ of \cite{YoudinGoodman05}, thought to play a critical role in planet formation by concentrating the dust into gravitationally bound clumps that may collapse into planetesimals \citep{JohansenYoudin07, Johansen+09}. More recently, another drift instability has been observed by \cite{Lambrechts+16} in their simulations of dust sedimentation.

Though such drag-and-drift instabilities have been known for decades, researchers still struggle to understand the detailed physics behind their onset. This is not surprising, considering the difficulties inherent in dealing with two coupled components (gas and dust). Nonetheless, it is a deficiency of the theory: certainly, it is hard to ask the right questions of a poorly understood phenomenon, and perhaps this motivates the large number of studies guessing what physical effects, ignored in \cite{YoudinGoodman05}, strongly impact on the \SI\ (see, \textit{e.g.} \citealt{Lin21} and references therein). Moreover, without sound physical understanding, it is more challenging to interpret numerical simulations and, in particular, disentangle physical from numerical effects.

In the particular case of the \SI, because it is so important, there have been some efforts to explain how the instability works \citep{YoudinJohansen07, Jacquet+11, LinYoudin17}. But the most important progress has been due to Squire \& Hopkins. In \cite{SquireHopkins18b} -- hereafter \citetalias{SquireHopkins18b}: they found that many drag-and-drift instabilities found in different systems are part of the same class of instabilities, which they called `resonant drag instabilities' or \RDIs. They showed that in almost every system where there exists a gas wave whose phase velocity is equal to the dust's drift velocity, there also exists a fast instability. Then in \cite{SquireHopkins18a} they found that the \SI\ is an \RDI\ arising from `epicyclic oscillations',\footnote{\cite{Zhuravlev19} clarified that the \SI\ is an \RDI\ arising from inertial waves rather than epicyclic oscillations.} and in \cite{SquireHopkins20} -- hereafter \citetalias{SquireHopkins20} -- they tried to provide a physical picture for this `epicyclic' \RDI. Unfortunately, their explanation remains partial: it does not explain the role played by the resonance, it is not accompanied by mathematical justifications, and it does not address \RDIs\ generally -- only the \SI.

Our goal in the present paper is to make a first step towards explaining \RDIs\ in general terms. For this first paper, we consider the simplest \RDI, the acoustic one, because a simple example is often instrumental in understanding the general case. Indeed in our next paper, we will see that the picture built here for the acoustic \RDI\ already contains all the concepts one needs to explain the general \RDI.  We also provide a new mathematical framework, which is equivalent to the linear algebra approach of \citetalias{SquireHopkins18b}, but trades efficiency for transparency. This approach clearly outlines the physical processes behind the acoustic \RDI.

We show that the acoustic \RDI\ is founded on two simple processes, which connect to each other in an unstable feedback loop. In the first mechanism, a sound wave concentrates dust because its compressive flows drag the dust, forcing it to accumulate in the regions of flow convergence. In the second mechanism, a dust density perturbation exerts a significant back-reaction on the gas; this force has just the right pattern to accelerate the flows of the original sound wave, thereby strengthening it. This stronger wave can then concentrate dust further, leading to larger density perturbations, which make the wave even stronger, and so on, and so forth... Similar ideas were explored by \citetalias{SquireHopkins20}, because this feedback loop is at the heart of every \RDI. The only difference between the physical pictures of two given \RDIs\ is their first mechanism: each particular gas wave concentrates dust in its own way.

The structure of the paper is as follows. We start in \S\ref{sec:prelim} with a short summary of the physics of the acoustic \RDI. This section will help the first-time reader form an intuition for what is happening, before they delve into the mathematical justification. Then starts the main part of the paper. First in \S\ref{sec:governing_equations} we present the physical system and the equations governing it, then in \S\ref{sec:1D} we explain in detail how the acoustic \RDI\ develops in the 1D version of this system. Finally, we briefly explain in \S\ref{sec:2D} why the 3D instability grows in exactly the same way as the 1D instability, and we conclude in \S\ref{sec:conclusion}. The appendices show how our novel intuition for the acoustic \RDI\ makes it easy to investigate how the instability reacts to non-ideal conditions such as detuning (\S\ref{sec:detuning}), extreme wavelengths (\S\ref{sec:small_and_large_scale}) or variations in the stopping time (\S\ref{sec:variable_stopping_times}).

\section{Outline of the physics of the acoustic RDI}
\label{sec:prelim}

In this section we sketch out the physical basis of the instability, using mainly non-mathematical arguments. Our goal is to develop a physical understanding of its underlying mechanism, which will help motivate and clarify the mathematical formalism presented later.

Our exposition concerns the one-dimensional case, the simplest possible version of the instability and thus the easiest to understand. The 1D case includes all the ingredients of the 3D version, and thus is a natural starting point. In particular, it illustrates clearly the two key processes that combine to make exponential growth: concentration of dust by a sound wave (the `forward' action), and the strengthening of that sound wave by said concentration (the `backward' reaction). Each process is treated separately, then discussed in combination. Finally, the discussion is generalised to 3D, to detuned modes, and to a different drag regime.

\subsection{Model set-up}
\label{sub:intuition_setup}

 Before beginning, we define our setup. Consider an isothermal gas in 1D, with sound speed ${ \cs }$, permeated by dust modelled as a pressure-less and uniform fluid. Suppose the dust drifts relative to the gas at some velocity ${ v }$, and the gas and dust act upon each other via Stokes drag.

The gas supports two linear sound waves, propagating either to the right or to the left with speed ${ \cs }$. Now consider the special case ${ v = \cs }$, \textit{i.e.} when the dust drifts at the same speed as one of the sound waves. The dust is `in resonance' with that wave.

\subsection{Forcing of a dust density perturbation by a sound wave}
\label{sub:intuition_forward}

\begin{figure}
    \centering
    \includegraphics[width = 0.95 \linewidth]{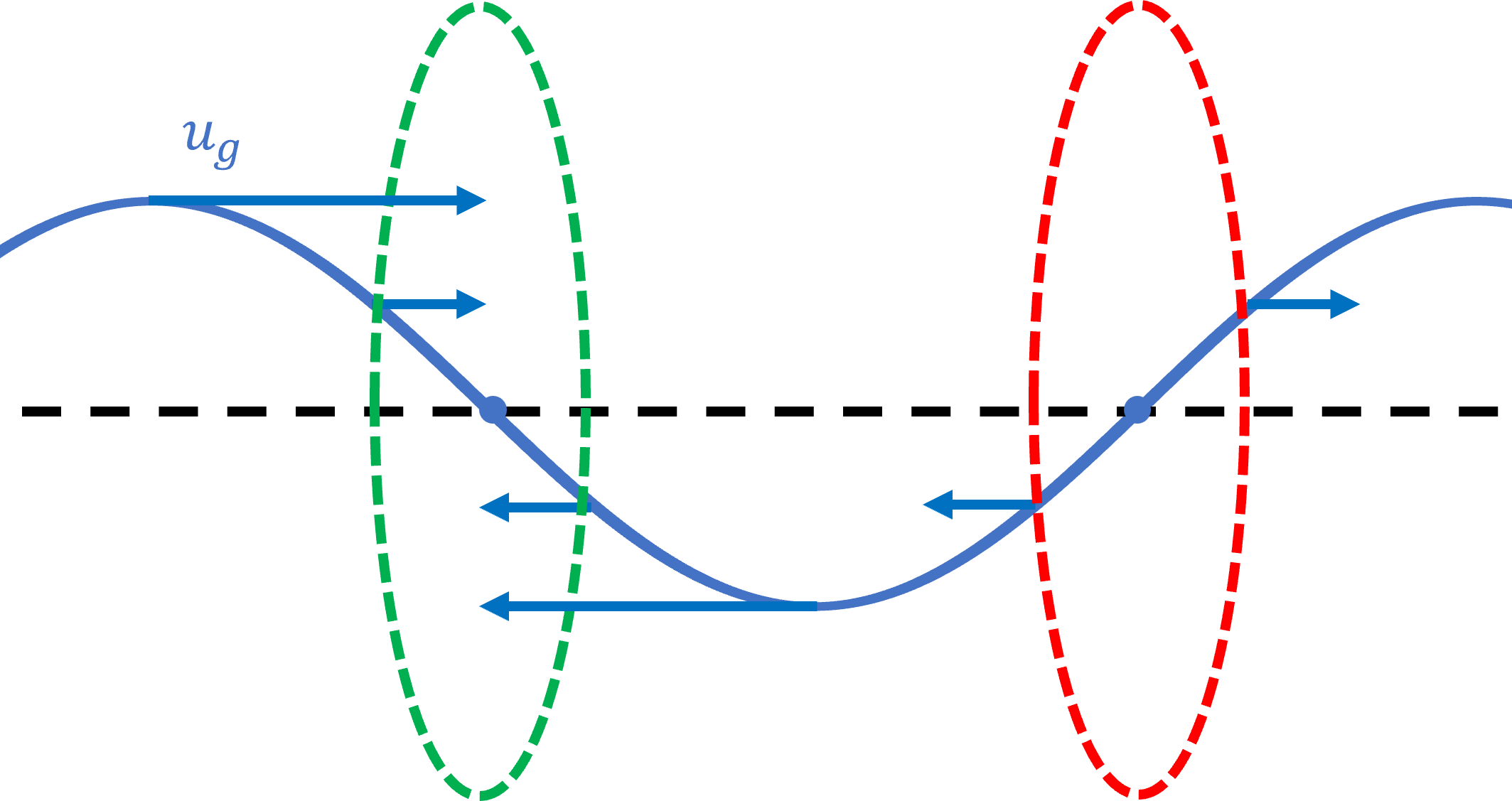}
    \caption{Schematic drawing of the forward action by which a sound wave is able to concentrate dust (\textit{cf.}~\S\ref{sub:intuition_forward}). The gas velocity's spatial profile ${ \ug }$ is represented both by the blue sine wave and the blue arrows; the dashed black line represents zero velocity. The dashed green/red oval illustrates a region of converging/diverging drag, and hence dust concentration/evacuation.}
    \label{fig:Dust_accumulation_acoustic}
\end{figure}

The spatial profile of the gas velocity due to the sound wave, ${ \ug }$, is sketched out in Fig.~\ref{fig:Dust_accumulation_acoustic}
(blue sine wave and arrows). It comprises a characteristic sequence of converging and diverging flows. At resonance, this flow pattern appears stationary in a frame moving with the background dust drift $v$. 

Now, because of drag, the initially uniform dust will experience a force proportional to the blue arrows, and will start following them. As a consequence, in regions of converging gas flow (dashed green oval), the drag will concentrate the dust. And because we are at resonance, dust does not drift out of the green oval, so the dust density there will keep growing indefinitely as more and more dust is brought in by the drag. Conversely, in regions of diverging gas flow (dashed red oval), the dust will be evacuated indefinitely. This can be formalised as an algebraic instability, whose `algebraic growth rate' is proportional to the gas' velocity divergence. 

The growing peaks and troughs in the dust density pattern are located at the nodes of the sound wave's velocity pattern. Thus, the dust perturbation will grow with a phase shift of ${ \pi / 2 }$ relative to the gas wave.

\subsection{Forcing of a sound wave by a dust density perturbation}
\label{sub:intuition_backward}

\begin{figure}
    \centering
    \includegraphics[width = 0.95 \linewidth]{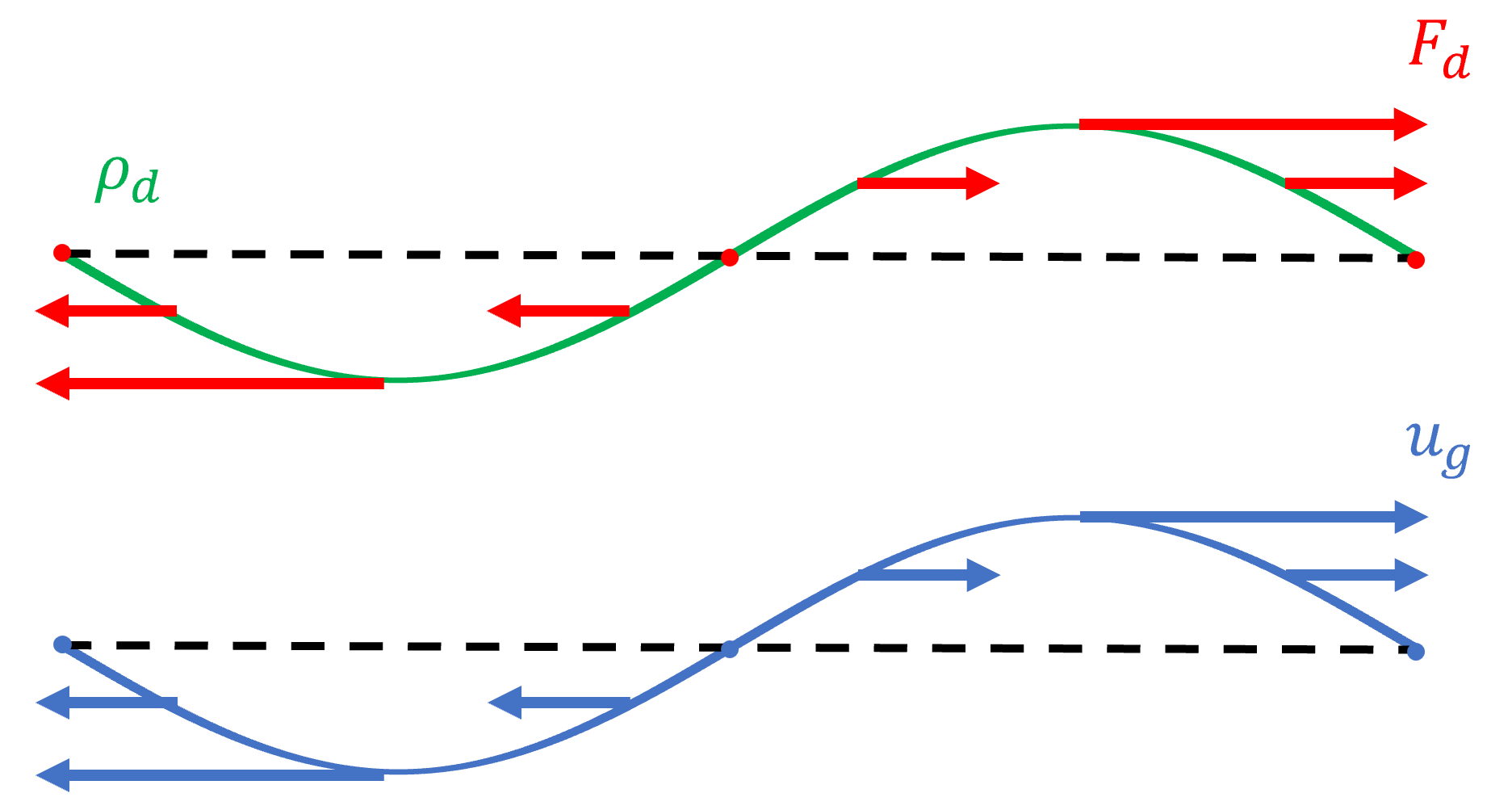}
    \caption{Schematic drawing of the backward reaction by which a sound wave is amplified by a dust density perturbation (\textit{cf.}~\S\ref{sub:intuition_backward}). In the top panel, the dust density perturbation is given in green, and the resulting drag force by the red arrows. In the bottom panel, the sound wave velocity is represented by blue arrows and the blue sine wave. The dashed black lines represent zero density/drag in the top panel and zero velocity in the bottom panel.}
    \label{fig:Impact_of_dust_feedback_acoustic}
\end{figure}

If the dust concentrates to sufficient levels, then we might expect it to react back on the gas via drag. To simplify how that may work, consider a sinusoidal dust density perturbation of arbitrary provenance \emph{in phase} with a pre-existing sound wave of negligible amplitude.\footnote{Note that if the gas is initially unperturbed, then the mechanism below can force a sound wave. We only assume a pre-existing sound wave to make the explanation concise.} Our example dust density perturbation ${ \rhod }$ is represented in green in the top panel of Fig.~\ref{fig:Impact_of_dust_feedback_acoustic}, while the sound wave is depicted in the bottom panel, with the blue arrows denoting the gas velocity field ${ \ug }$. Suppose also that the dust has no velocity perturbation.

The drag force exerted by the dust on the gas is more complicated than that exerted by the gas on the dust because it depends on the relative densities of dust and gas (see Eq.~\ref{eq:Navier_Stokes_acoustic_momentum_gas}). In particular, dust over-densities and under-densities perturb the drag force away from its equilibrium value. We represent the excess drag by the red arrows in the top panel of Fig.~\ref{fig:Impact_of_dust_feedback_acoustic}. This force has just the right spatial profile to accelerate the sound wave's flow everywhere, thereby strengthening the wave. Because the dust perturbation drifts with the gas wave (i.e. it is resonant, see \S\ref{sub:intuition_setup}), this forcing will remain stationary with respect to the wave. As a consequence, the wave will grow indefinitely, in an algebraic fashion (as earlier).

\subsection{Combining the two growth processes}
\label{sub:intuition_combine}

An obvious discrepancy in our picture so far is that the dust perturbation generated by a sound wave (\S\ref{sub:intuition_forward}) is ${ \pi / 2 }$ out of phase with the wave, whereas the sound wave preferentially forced by the dust perturbation (\S\ref{sub:intuition_backward}) is perfectly in phase. Thus the two processes cannot be joined in a straightforward way. In fact, as we shall show later, the exponentially growing \RDI\ must operate via a compromise: its dust density and gas velocity patterns are out of phase by ${ \pi / 4 }$, precisely the arithmetic mean of the two limiting processes. Both dust concentration and sound wave forcing are now inefficient, but can still work well enough to create a positive feedback loop and hence exponential instability.  In fact, the growth rate of the exponential instability is the geometric mean of the 'algebraic growth rates' of its forward action and backward reaction.

\subsection{Generalisations}
\label{sub:intuition_generalisations}

The setup in which we have explained the acoustic \RDI\ so far is quite limited. There are three generalisations of this setup that we think are crucial to cover, because they greatly increase the prevalence of the acoustic \RDI.

\subsubsection{The instability in three dimensions}
\label{ssub:intuition_3D}

Generalising the process to 3D is relatively straightforward. If ${ \mathbf{\hat{k}} }$ denotes the direction in which the sound wave propagates and ${ \bv }$ is the dust's drift velocity, we decompose ${ \bv }$ into components parallel and perpendicular to ${ \mathbf{\hat{k}} }$, i.e. ${ \bv = \vpara \, \mathbf{\hat{k}} + \bvperp }$ where ${ \bvperp \bcdot \mathbf{\hat{k}}=0 }$. The 3D resonance condition becomes simply ${ \vpara = \cs }$, which is the same as the 1D resonance condition, but with $v$ replaced by its component parallel to the wave's phase velocity. 

\begin{figure}
    \centering
    \includegraphics[width = 0.7 \linewidth]{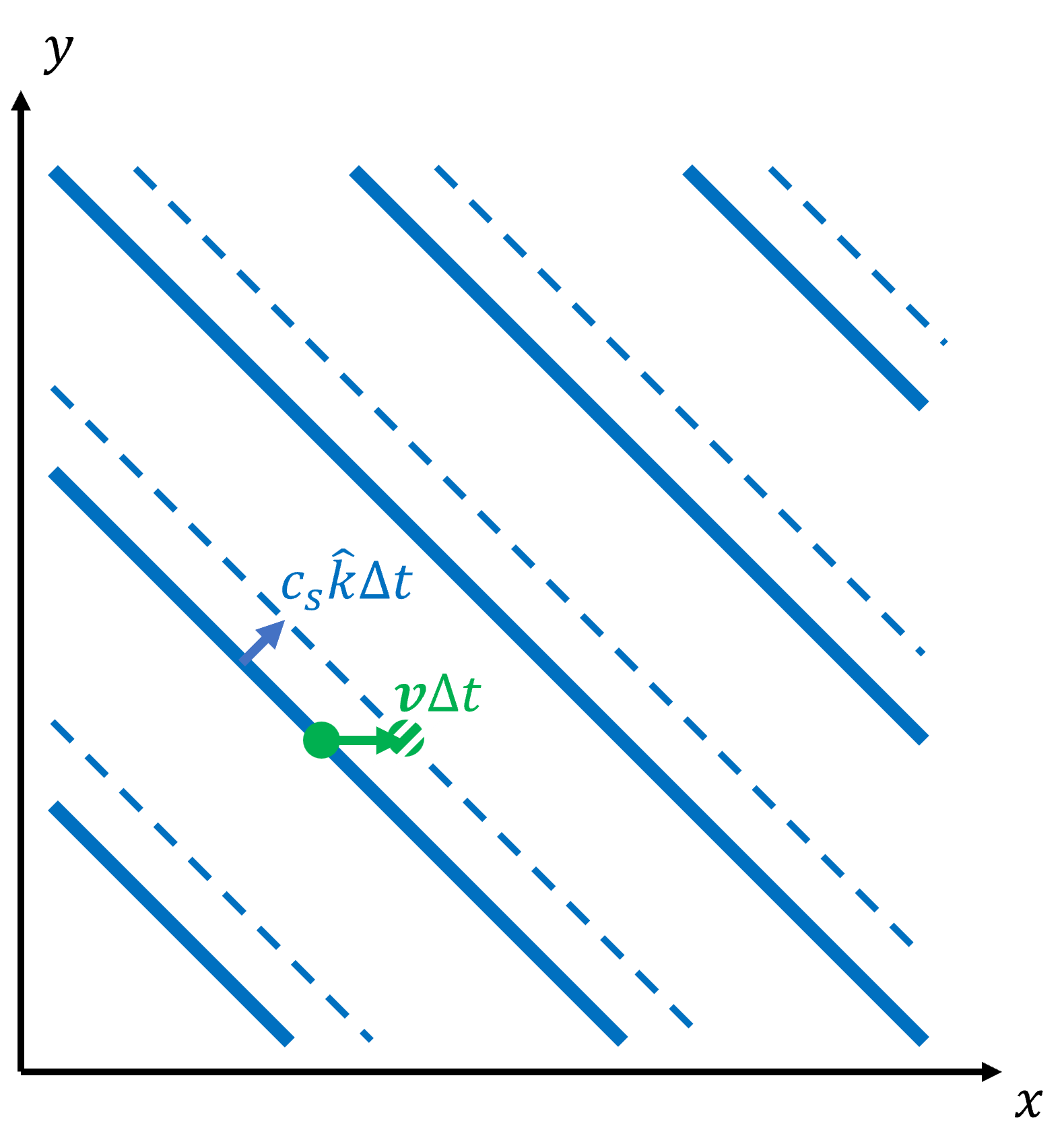}
    \caption{Schematic drawing of the multi-dimensional resonance condition (\textit{cf.} \S\ref{sub:intuition_generalisations}). The solid blue lines represent the crests of a sound wave at time ${ t }$, and the solid green dot a dust particle that sits on a wave crest. The dashed blue lines and the dashed green circle represent the situation at a time $\Delta t$ later. If the resonance condition holds, the dust particle remains on the same wave crest.
    }
    \label{fig:Resonance_condition}
\end{figure}

The 3D resonance condition is illustrated in Fig.~\ref{fig:Resonance_condition}, which shows the peaks of a sound wave (blue) and also the location of a dust particle (green) at two times, $\Delta t$ apart (solid and dashed), with the dust drift $\bv$ in the $x$-direction. At resonance, the figure tells us that drifting dust starting on a peak (solid green circle) will remain on that peak at any later time (dashed green circle).
Essentially, the resonance condition demands that the dust only drifts within waveplanes. In the forward action this ensures that a dust particle stays within the region of dust accumulation. In the backward reaction, it ensures that the drag force exerted by the dust on the gas remains \textit{in phase} with the sound wave.

Crucially, there is no constraint on ${ \bvperp }$. This extra degree of freedom in 3D makes the acoustic \RDI\ much more viable. In the 1D case the instability exists only when the dust drift velocity is near sonic, whereas in 3D it can be supersonic.

\subsubsection{The impact of detuning}
\label{ssub:intuition_detuning}

So far our explanation for the instability required perfect resonance, but significant exponential growth can still be achieved when the system is off resonance. The degree of detuning tolerated by the \RDI\ scales with the square root of the dust-to-gas ratio (see \S\ref{sec:detuning} for a proof). So even in a dust-poor medium whose gas-to-dust ratio is in the thousands, there are several percent of leeway on the resonance condition. This again makes the acoustic \RDI\ more pervasive.

\subsubsection{In the Epstein regime of drag}
\label{ssub:intuition_varying_stopping_time}

In the Stokes regime, the drag coefficient of the dust is a constant. But small dust grains are in the Epstein regime, where the drag coefficient depends on the gas density and the relative velocity between dust and gas \citep{ChiangYoudin10}. 

The main effect of these dependencies is to reduce the efficiency of the mechanism of \S\ref{sub:intuition_forward}, by reducing the tendency of the dust to follow the converging gas flows. As a consequence, the growth rate of the instability tends to be lower in the Epstein regime than in the Stokes regime, though instability is certainly not quenched. Note that all these statements are made quantitative in \S\ref{sec:variable_stopping_times}.

\section{Presentation of the physical system}
\label{sec:governing_equations}

\subsection{Governing equations}
\label{sub:governing_equations}

Consider a mixture of gas and dust, modelled as a two-fluid system. The gas is assumed to be isothermal and inviscid, and it is described by its density ${ \rhog }$, its speed ${ \bug }$, and its pressure ${ P }$. The dust is assumed to be pressure-less, and is described by its density ${ \rhod }$ and velocity ${ \bud }$. The two fluids are coupled by a drag force from the gas onto the dust, and its back-reaction from the dust onto the gas. This is summed up in the equations
\begin{subequations}
    \label{eq:Navier-Stokes_acoustic}
    \begin{align}
        \pD t \rhog + \bug \bcdot \grad \rhog \, &= - \rhog \grad \bcdot \bug , \label{eq:Navier_Stokes_acoustic_continuity_gas} \\
        \pD t \bug + \bug \bcdot \grad \bug \, &= - \frac{1}{\rhog} \grad P + \frac{\dtg}{\tau} (\bud - \bug) + \mathbf{f_{g}}, \label{eq:Navier_Stokes_acoustic_momentum_gas} \\
        \pD t \rhod + \bud \bcdot \grad \rhod \, &= - \rhod \grad \bcdot \bud , \label{eq:Navier_Stokes_acoustic_continuity_dust} \\
        \pD t \bud + \bud \bcdot \grad \bud \, &= - \frac{1}{\tau} (\bud - \bug) + \mathbf{f_{d}} , \label{eq:Navier_Stokes_acoustic_momentum_dust} \\
        P \, &= \cs^{2} \, \rhog , \label{eq:Navier_Stokes_acoustic_isothermal_gas}
    \end{align}
\end{subequations}
where ${ \dtg = \rhod / \rhog }$ is the dust-to-gas ratio, ${ \tau }$ is the dust's stopping time, and ${ \cs }$ is the gas' constant sound speed. ${ \mathbf{f_{g}} }$ and ${ \mathbf{f_{d}} }$ are accelerations due to undisclosed external forces acting respectively on the gas and the dust. These forces need not be the same for the two species. For instance, in a molecular cloud, only the dust feels the radiation pressure, the photoelectric force and the photodesorption force due to the hard radiation field from nearby massive stars \citep{Weingartner01}. However, we do assume that ${ \mathbf{f_{g}} }$ and ${ \mathbf{f_{d}} }$ are uniform in space and constant in time.

This physical system is almost identical to that of \citetalias{HopkinsSquire18a}, but with two simplifications. First, we consider an isothermal gas with a fixed sound speed, rather than a barotropic gas, but this has no impact on the physics or mathematics of the linear instability. Second, we assume that the dust's stopping time is a fixed parameter, rather than a variable. This assumption is valid in the isothermal \& Stokes regime, but not in the Epstein regime. We shall relax it in \S\ref{sec:variable_stopping_times}. But for now, we favour the simpler model because it enhances the clarity of our explanation for the acoustic \RDI.

\subsection{Background equilibrium flow}
\label{sub:equilibrium}

The next step is to describe the equilibrium flow upon which infinitesimal perturbations will eventually grow. We shall denote background quantities with an overline ${ \overline{\phantom{a}} }$.\\

The simplest solution to Eq.~\eqref{eq:Navier-Stokes_acoustic} is obtained by assuming constant and homogeneous gas and dust densities ${ \rhogo }$ and ${ \rhodo }$, and uniform velocities ${ \bugo (t) }$ and ${ \budo (t) }$. These assumptions lead to 
\begin{equation}
    \label{eq:background_acoustic}
    \bugo (t) = \mathbf{a} t \,\,\,\,\,\,\,\text{ and }\,\,\,\,\,\,\, \budo (t) = \mathbf{a} t + \bv .
\end{equation}
where
\begin{equation}
    \mathbf{a} =  \frac{\mathbf{f_{g}} + \dtgo \, \mathbf{f_{d}}}{1 + \dtgo} 
     \qquad \text{and}\qquad 
     \bv = \tau \frac{\mathbf{f_{d}} - \mathbf{f_{g}}}{1 + \dtgo} .
    \label{eq:background_drift_acoustic}
    \end{equation}
$\mathbf{a}$ is an acceleration common to gas and dust, and $\bv$ is the drift of the dust relative to the gas.

As \citetalias{HopkinsSquire18a} point out, it is preferable to work in a new reference frame ${ \mathcal{R}_{acc.} }$ with the same axes but whose origin is moving with the gas. In this reference frame, the background flow simplifies to
\begin{equation}
    \label{eq:background_acoustic_v2}
    \overline{\mathbf{u}}_{g, acc.} = \mathbf{0} \,\,\,\,\,\,\,\text{ and }\,\,\,\,\,\,\, \overline{\mathbf{u}}_{d, acc.} = \bv .
\end{equation}
In compensation, because ${ \mathcal{R}_{acc.} }$ is accelerating, it is non-Galilean and we must add a fictitious force ${ \mathbf{F_{fic.}} = - \mathbf{a} }$ to the r.h.s. of the momentum equations (\ref{eq:Navier_Stokes_acoustic_momentum_gas}, \ref{eq:Navier_Stokes_acoustic_momentum_dust}). But because this force is uniform in space and constant in time, it will not appear in the perturbation equations. So all in all, switching to ${ \mathcal{R}_{acc.} }$ is a net simplification. We shall work exclusively in this reference frame from now on, and shall drop the subscript ${ _{acc.} }$ to improve readability.\footnote{Alternatively, one might prefer to explicitly change coordinates: ${ (\bx, t) \mapsto (\bx' = \bx - \frac{1}{2}\mathbf{a} t^{2}, t' = t) }$ with the velocities boosted by ${ - \mathbf{a} t }$.}

\subsection{Linearised equations}
\label{sub:linearised_equations}

\enlargethispage{\baselineskip}
Let us now disturb the background flow, ${ \overline{f} }$ with a perturbation, ${ f' }$, which we assume to be small, ${ f' \ll \overline{f} }$. The total flow is then ${ f = \overline{f} + f' }$, where ${ f }$ denotes any unknown (pressure, density, velocity). At the linear order, this leaves the perturbation equations
\begin{subequations}
    \label{eq:perturbation_equations_acoustic}
    \begin{align}
        & \pD t \rgi = - \grad \bcdot \bugi , \label{eq:perturbation_equations_acoustic_continuity_gas} \\
        & \pD t \bugi =  - \cs^{2} \, \grad \rgi + \frac{\dtgo}{\tau} (\budi - \bugi) + \frac{\dtgo}{\tau} \bv \, (\rdi - \rgi) , \label{eq:perturbation_equations_acoustic_momentum_gas} \\
        & \pD t \rdi + \bv \bcdot \grad \rdi = - \grad \bcdot \budi , \label{eq:perturbation_equations_acoustic_continuity_dust} \\
        & \pD t \budi + \bv \bcdot \grad \budi =  - \frac{1}{\tau} (\budi - \bugi) , \label{eq:perturbation_equations_acoustic_momentum_dust}
    \end{align}
\end{subequations}
where ${ \varrho_{i}' = \rho_{i}'/\overline{\rho_{i}}} $ is the relative perturbation in density, either for the gas or for the dust.

\section{One-dimensional case}
\label{sec:1D}

The 1D case simplifies the physics of the \RDI\ and keeps only what is essential to drive the instability. In that sense, it is an instructive case to consider.

We start in \S\ref{sub:dispersion_relation} by considering the 1D dispersion relation, which shows instability is possible when the gas and dust are at resonance. This is the 1D acoustic \RDI. The ensuing subsections aim at providing a physical picture of this instability. First in \S\ref{sub:1D_forward} we explain the forward mechanism, by which a sound wave can concentrate dust. Then in \S\ref{sub:1D_backward} we uncover the backward mechanism, by which a dust density perturbation can strengthen sound waves. Finally in \S\ref{sub:1D_combine} we demonstrate how these two mechanisms feed into each other at resonance, creating an unstable feedback loop, leading to the 1D acoustic \RDI.

\subsection{Dispersion relation, resonance and instability}
\label{sub:dispersion_relation}

The goal of this first subsection is to extract as much information as possible from the dispersion relation. This will allow us to understand why the instability is called `resonant', and to estimate its growth rate.

First of all, let us establish the 1D dispersion relation. To do this, we look for modal solutions to Eqs.~\eqref{eq:perturbation_equations_acoustic}, \textit{i.e.} solutions that can be decomposed into full Fourier modes of the form ${ f(t, x) = \widehat{f} \, \e^{i (k x - \omega t)} }$. Such solutions exist if ${ \omega }$ and ${ k }$ satisfy
\begin{equation}
    \label{eq:dispersion_relation_acoustic_1D}
    (\omega^{2} - \cs^{2} k^{2}) (\omega - v k) (\omega - v k + i / \tau) + \dtg \, P(\omega, v k) = 0 ,
\end{equation}
where
\begin{equation}
    \label{eq:P_acoustic}
    P(\omega, v k) = \frac{i}{\tau} \! \left[ \omega^{3} \!\! - \! v k \omega^{2} \! - \! v^{2} k^{2} \omega \! + \! v^{3} k^{3} \! - \! \frac{i v^{2} k^{2}}{\tau}  \right] \! 
\end{equation}
is a multi-variate polynomial.

This can be reframed using only three dimensionless numbers: the dust-to-gas ratio $\dtg$, the Mach number of the drift ${M = v / \cs }$ and the dimensionless wave-number ${K = \cs k \tau }$. The unusual thing is that there is no Stokes number, even though drag plays a key role in the dynamics of the instability. This reflects the fact that there is a single timescale in the system: the stopping time of the dust.

\subsubsection{Resonance}
\label{ssub:resonance}

When ${ \dtg = 0 }$, the dispersion relation~\eqref{eq:dispersion_relation_acoustic_1D} describes two sound waves (${ \omega = \pm \cs k }$), a dust advective mode (${ \omega = v k }$), and a damped dust mode (${ \omega = v k - i / \tau }$). When the dust drift is transonic, ${ v \approx \cs }$, the right-propagating sound wave and the dust advective mode possess similar frequencies. Let us investigate what happens then.

${ \dtg = 0 }$ marks the regime of test particles, \textit{i.e.} the regime in which the dust is too sparse to exert a significant back-reaction on the gas. As such, the gas behaves as in isolation. So, we can select the right-propagating sound wave of frequency ${ \omega = \cs k }$ and inspect its effect on the dust perturbation. We find
\begin{equation}
    \label{eq:explain_resonance_acoustic}
    \hrd = \frac{i \cs \, \hrg / \tau}{(\cs  - v ) (\cs k - v k + i / \tau)} ,
\end{equation}
which reveals that when ${ v \approx \cs }$, the dust responds extremely strongly to being forced by a sound wave. This justifies the use of the term `resonance' in `resonant drag instability'. 

Moreover, at perfect resonance, ${ v = \cs }$, there is a singularity and the dust's response explodes. This indicates that the modal ansatz is inappropriate at resonance.\footnote{Ultimately, this is because the right-propagating sound wave and the dust advective mode collide not only in frequency but also in eigenmode. Therefore, if we count the modal solutions, we find that at resonance we have one less than usual. The missing solution must be non-modal. For more details on mode collisions, see \cite{Zhuravlev19} and references therein.} We need to consider non-modal solutions to Eqs.~\eqref{eq:perturbation_equations_acoustic}, \textit{i.e.} solutions that are Fourier decomposed only in space, so that ${ f(t, \bx) = \widetilde{f}(t) \, \e^{i \bk \cdot \bx} }$. Note that we use a hat ${ \, \widehat{} \, }$ to denote the amplitude of a full Fourier mode, but a tilde ${ \, \widetilde{} \, }$ for the time-dependent amplitude of a spatial Fourier mode.

\subsubsection{Instability}
\label{ssub:instability_from_dispersion_relation}

Now if we stay at resonance but leave the regime of test particles, so that $\mu\neq 0$, this singularity disappears. The colliding modes are slightly modified by the back-reaction of the dust onto the gas, thereby avoiding each other. 

\citetalias{SquireHopkins18b} showed that these modifications are of order ${ \mathcal{O} (\dtg^{1/2}) }$. So, we perform an expansion in powers of ${ \dtg^{1/2} }$, which writes ${ \omega = \omegaO + \sqrt{\dtg} \, \omegaI + ... }$ with ${ \omegaO = v k = \cs k }$. When the dimensionless wave-number $K$ is of order unity, we obtain
\begin{equation}
    \label{eq:growth_rate_from_dispersion_relation_acoustic}
    \omegaI^{2} = \frac{i \tau \, P(\cs k, \cs k)}{2 \cs k} = \frac{i \cs k}{2 \tau} ,
\end{equation}
and thus
\begin{equation}
    \label{eq:growth_rate_from_dispersion_relation_acoustic_v2}
    \gamma = \mathfrak{Im} (\omega) = \frac{\dtg^{1/2}}{2} \sqrt{\frac{ \cs k}{\tau}} + \mathcal{O} (\dtg^{1}),
\end{equation}
which shows that the back-reaction from the dust onto the gas causes an instability: the acoustic \RDI. This growth rate was first computed in Eq.~13 of \citetalias{SquireHopkins18b}. It also appears in Eqs.~15 and 19 of \citetalias{HopkinsSquire18a}, who call it the ``intermediate wavelength resonant quasi-drift mode".

This formula is only valid when the dimensionless wave-number ${ K = \cs k \tau }$ is of order unity. We relax this assumption in \S\ref{sec:small_and_large_scale} and find that there are three regimes: long, intermediate, and short wavelengths. On small scales, the pressure-less approximation may fail. But if it remains valid, the growth rate scales as ${ K^{1/3} }$, so there is a divergence as $K \!\! \to \! \infty$ but it is weaker than suggested by Eq.~\eqref{eq:growth_rate_from_dispersion_relation_acoustic_v2}. We propose a tentative explanation for this divergence at the end of \S\ref{sub:1D_combine}. On large scales, the instability is weaker than Eq.~\eqref{eq:growth_rate_from_dispersion_relation_acoustic_v2} suggests.

It also seems like the growth rate scales with ${ \tau^{-1/2} }$ and thus diverges in the limit of well-coupled dust. This would be counter-intuitive, and indeed is not what happens. Remember that the resonance condition imposes ${ \cs = v }$, and that ${ v \propto \tau \, (f_{g} - f_{d}) }$. Therefore, if the external forces $f_{g}$ and ${f_{d}}$ are fixed, the ${ \cs \, \tau^{-1} }$ term in Eq.~\eqref{eq:growth_rate_from_dispersion_relation_acoustic_v2} is actually a constant. Or, if ${ \cs }$ is fixed, we need very strong external forces to sustain a fast enough drift. These forces are ultimately the source of free energy for the acoustic \RDI, so it makes sense that if they are strong then the instability is strong. \\

\enlargethispage{1 \baselineskip}
The goal of the present paper is to explain the physics of this instability. The dispersion relation gave us the growth rate, and highlighted the role of a resonance between two waves. But we still need to understand how those two waves interact, and why it leads to an instability.

\subsection{Forward action: accumulation of dust by a sound wave}
\label{sub:1D_forward}

Let us start by capitalizing on the insights gained from Eq.~\eqref{eq:explain_resonance_acoustic} and see what we can learn from a non-modal analysis in the regime of test particles, ${ \dtg = 0 }$. This investigation shall uncover the first constituent mechanism of the acoustic \RDI, by which a sound wave can concentrate dust (see \ref{sub:intuition_forward}).

\subsubsection{There is an algebraic instability of dust forced by a sound wave}
\label{ssub:1D_forward_block_1}

Since the dust is insufficiently dense to affect the gas, the sound wave should always exist in this regime, even at resonance. Simply, it will exert a forcing on the dust, leading to a non-modal response. Therefore, let us assume a modal sound wave ${ f_{g} (t, x) = \widehat{f_{g}} \, \e^{i (k x - \cs k t)} }$ for the gas, and a non-modal answer ${ f_{d} (t, x) = \widetilde{f_{d}}(t) \, \e^{i k x} }$ from the dust. Here $\widetilde{f_{d}}(t)$ is a time-dependent amplitude to be determined.

The gas equations (\ref{eq:perturbation_equations_acoustic_continuity_gas}, \ref{eq:perturbation_equations_acoustic_momentum_gas}) are satisfied by the sound wave, and the dust equations (\ref{eq:perturbation_equations_acoustic_continuity_dust}, \ref{eq:perturbation_equations_acoustic_momentum_dust}) become
\begin{subequations}
    \label{eq:IVP_dust_acoustic}
    \begin{align}
        & \D t \trd + i \, v k \, \trd = - i \, k \tud , \label{eq:IVP_dust_acoustic_continuity} \\
        & \D t \tud + i \, v k \, \tud =  - \frac{1}{\tau} (\tud - \hug \, \e^{-i \, \cs k \, t}) , \label{eq:IVP_dust_acoustic_momentum}
    \end{align}
\end{subequations}
which leads to the forced second-order \ODE
\begin{equation}
    \label{eq:forcing_in_kernel_of_operator_dust_acoustic}
    \bigg( \D t + i \, vk + \frac{1}{\tau} \bigg) \, \bigg( \D t + i \, vk \bigg) \, \trd = - \frac{i}{\tau} \, k \hug \, \e^{-i \, \cs k \, t} .
\end{equation}
This form of the \textit{l.h.s.} makes it clear that at resonance, ${ v = \cs }$, the forcing frequency coincides with one of the dust's natural frequencies. As a consequence, the particular solution to Eq.~\eqref{eq:forcing_in_kernel_of_operator_dust_acoustic} is proportional to ${ t \, \e^{-i \, \cs k \, t} }$, which reveals algebraic growth with time. This is the algebraic instability reported in \S\ref{sub:intuition_forward}. The following subsections elaborate on its mathematical and physical features.

It should be stressed that this algebraic instability exists solely because we idealised the physics. Any non-ideal effect (in particular any detuning), however slight, would kill the algebraic instability. Fortunately, the algebraic instability is not the \RDI, which is exponential and very robust. The algebraic instability is useful to understand because it is a salient manifestation of one of the two key processes behind the acoustic \RDI: the process by which the dust density increases greatly when forced by a sound wave. And this process is robust. For instance, if we were slightly off-resonance, Eq.~\eqref{eq:forcing_in_kernel_of_operator_dust_acoustic} indicates that we would still see a transient regime of growth in dust density after we turn on the forcing, and a large final dust density.

\subsubsection{The dust density grows because the sound wave contains convergent flows and the dust follows them}
\label{ssub:1D_forward_block_2}

In \S\ref{sub:intuition_forward} we gave a heuristic account for this dust concentration process. To explain where our intuition came from and to show that it is backed by mathematics, let us solve Eqs.~\eqref{eq:IVP_dust_acoustic}. \\

Since at resonance the wave and the dust have the same velocity, it is much more convenient to work in the co-moving reference frame. Equations~\eqref{eq:IVP_dust_acoustic} simplify to
\begin{subequations}
    \label{eq:IVP_dust_acoustic_v2}
    \begin{gather}
        \D t \trd = - i \, k \tud   \quad  \text{and} \quad  \D t \tud =  - \frac{1}{\tau} (\tud - \hug) . \tag{\theequation a-b}
    \end{gather}
\end{subequations}
The second equation is independent of the first. Its general solution is
\begin{equation}
    \label{eq:solution_dust_velocity_over_time_acoustic}
    \tud (t) = \hug + c_{1} \, \e^{-t / \tau} ,
\end{equation}
where ${ c_{1} }$ is an undetermined integration constant. Clearly, the second term goes to zero over time, leaving only the first term. This shows that whatever the initial conditions, after just a few stopping times, the dust follows the gas flow. Therefore, on long timescales, Eq.~(\ref{eq:IVP_dust_acoustic_v2}a) becomes
\begin{equation}
    \label{eq:algebraic_growth_rate_dust_acoustic}
    \D t \trd =  - i \, k \hug ,
\end{equation}
which confirms that only the dust density grows, and that it grows because the flows of a sound wave are convergent (${ i \, k \hug }$ translate to ${ \grad \bcdot \bug }$ out of Fourier space).

These last two equations justify the claims of \S\ref{sub:intuition_forward} and Fig.~\ref{fig:Dust_accumulation_acoustic}. The gas velocities arising from a sound wave are made of convergent and divergent flows. Because of drag, the initially uniform dust is forced to follow these flows, and converges to the region where the gas flows are compressive, forming a dust clump.

As shown by the green oval of Fig.~\ref{fig:Dust_accumulation_acoustic}, the flows of a sound wave converge to the velocity node that is just to the right of a velocity bump. This implies that the dust clump forms at a phase-shit of ${ - \pi / 2 }$ with respect to the sound wave, in agreement with the ${ -i }$ factor in Eq.~\eqref{eq:algebraic_growth_rate_dust_acoustic}. This ${ 90^{\circ} }$phase-shift is due to the concentration mechanism involving a divergence, \textit{i.e.} a single spatial derivative.

Finally, let us define the 'algebraic growth rate' of this forward action as
\begin{equation}
    \label{eq:algebraic_growth_rate_dust_acoustic_v2}
    \gamma_{1} = \frac{\D t \trd}{\trd (t= 0)} = \frac{-i \, k \hug}{\trd (t = 0)} .
\end{equation}

\subsubsection{Growth happens at resonance because the dust does not drift away from the velocity node}
\label{ssub:1D_forward_block_3}

Now that we understand the dust concentration mechanism, it is natural to wonder why it is only strong near resonance.
A clue comes from re-writing Eq.~(\ref{eq:IVP_dust_acoustic_v2}a) when we are out of resonance (and in the reference frame of the sound wave). It becomes
\begin{equation}
    \label{eq:IVP_dust_acoustic_out_of_resonance}
    \D t \trd + i k (v  - \cs ) \trd = - i \, k \tud .
\end{equation}
The only new thing brought by such a detuning is the advective term on the \textit{l.h.s.} It appears when the sound speed and the background drift velocity are mismatched, and dust particles drift relative to the sound wave. This demonstrates that the role of perfect resonance is to remove this drift, and the role of approximate resonance is to limit it.

Essentially, perfect resonance makes it impossible for dust that has arrived at the green oval of Fig.~\ref{fig:Dust_accumulation_acoustic} to escape it by drifting. As a consequence, the dust density grows forever in the oval as more and more dust is brought in by the gas flow. This is why algebraic growth in dust density is only possible at resonance. Close to resonance, dust only drifts very slowly out of the oval, so the residence time of dust in the oval is finite but very long, meaning that the oval can still contain a lot of dust. This is why a large increase in dust density is still possible close to resonance.

\subsection{Backward reaction: forcing of a sound wave by the dust}
\label{sub:1D_backward}

The preceding subsection focused on the first part of the acoustic \RDI, in which a sound wave can resonate with the dust flow and create a dust clump. Let us now leave the regime of test particles, \textit{i.e.} set ${ \dtg > 0 }$, and see how the back-reaction from the dust clump can excite sound waves. 

To best bring out this second constituent mechanism of the acoustic \RDI, we shall consider a somewhat idealised setup in which the dust is decoupled from the gas and only acts as a forcing. In a sense, this is symmetrical to what is done in \S\ref{sub:1D_forward}, where the gas was decoupled from the dust and acted as a forcing. This decoupling approach helps isolate each of the two mechanisms, thus bringing some clarity to our physical picture. Moreover, we will see in \S\ref{sub:1D_combine} that the decoupled equations studied in \S\ref{sub:1D_forward} and in the present subsection emerge naturally when we study the coupled equations in an asymptotic limit.

\subsubsection{There is another algebraic instability of gas forced by a dust density perturbation}
\label{ssub:1D_backward_block_1}

Let us assume a dust advective mode ${ f_{d} (t, x) = \widehat{f_{d}} \, \e^{i (k x - v k t)} }$ of amplitude ${ \hrd }$ that it is not impacted by any gas perturbation. This dust mode will exert a forcing on the gas, possibly leading to a non-modal response ${ f_{g} (t, x) = \widetilde{f_{g}}(t) \, \e^{i k x} }$. Let us also assume that initially, the gas perturbations are negligible compared to the dust density perturbation. The gas perturbations will gain amplitude over time, but it will take some time before they are significant compared to the dust perturbation. 
Therefore, we can safely neglect the drag terms proportional to gas density or velocity in Eq.~\eqref{eq:perturbation_equations_acoustic_momentum_gas}. 

This set of assumptions is simply a way to reduce the gas equations (\ref{eq:perturbation_equations_acoustic_continuity_gas}, \ref{eq:perturbation_equations_acoustic_momentum_gas}) to
\begin{subequations}
    \label{eq:IVP_gas_acoustic}
    \begin{align}
        & \D t \trg = - i \, k \tug , \label{eq:IVP_gas_acoustic_continuity} \\
        & \D t \tug =  - i \cs^{2} k \, \trg + \frac{\dtg v}{\tau} \hrd \, \e^{-i \, v k \, t} . \label{eq:IVP_gas_acoustic_momentum}
    \end{align}
\end{subequations}
We will see in \S\ref{sub:1D_combine} that this simpler set of equations emerges naturally from the study of the complete acoustic \RDI. The last term shows that the main effect of the dust clump is that its variations in dust density perturb the drag force exerted by the dust on the gas. In regions where there is more dust than at equilibrium, the gas feels a stronger push than at equilibrium. Conversely, in regions where there is less dust than at equilibrium, the gas feels a weaker push.

After some simple algebra, Eqs.~\eqref{eq:IVP_gas_acoustic} yield
\begin{equation}
    \label{eq:forcing_in_kernel_of_operator_gas_acoustic}
    \DD t \tug  + \cs^{2} k^{2} \tug = \frac{-i \dtg}{\tau} k v^{2} \, \hrd \, \e^{-i \, v k \, t} ,
\end{equation}
which is the equation of a forced harmonic oscillator. At resonance, ${ v = \cs }$, the forcing frequency is equal to the natural frequency of the oscillator. Hence, the particular solution to Eq.~\eqref{eq:forcing_in_kernel_of_operator_gas_acoustic} is proportional to ${ t \, \e^{-i \, v k \, t} }$, demonstrating the existence of another algebraic instability.

\subsubsection{It is a sound wave that grows}
\label{ssub:1D_backward_block_2}

To investigate what mode grows via this second instability, let us solve the \IVP\ presented by Eqs.~\eqref{eq:IVP_gas_acoustic}. At resonance, we find
\begin{subequations}
    \label{eq:solution_gas_over_time_acoustic}
    \begin{align}
        & \trg (t) = \,\,\,\,\,\,\,\,\,\, \lambda_{1} \e^{- i \, \cs k \, t} + \lambda_{2} \e^{+ i \, \cs k \, t} \,\,\,\, + \frac{1}{2} \frac{\dtg}{\tau} \hrd \, t \, \e^{- i \, \cs k \, t} , \label{eq:solution_gas_density_over_time_acoustic} \\
        & \tug (t) = \cs \left( \lambda_{1} \e^{- i \, \cs k \, t} - \lambda_{2} \e^{+ i \, \cs k \, t} \right) + \frac{\cs}{2} \frac{\dtg}{\tau} \hrd \, t \, \e^{- i \, \cs k \, t} , \label{eq:solution_gas_velocity_over_time_acoustic}
    \end{align}
\end{subequations}
where ${ \lambda_{1} }$ and ${ \lambda_{2} }$ are unimportant constants determined by the initial conditions. We are interested in the last terms, which do not depend on the initial conditions. As expected from Eq.~\eqref{eq:forcing_in_kernel_of_operator_gas_acoustic}, they grow linearly in time. What is significant, is that they exhibit the structure of a right-propagating sound wave. This confirms that the back-reaction from the dust clump is able to generate acoustic modes.

Eqs.~\eqref{eq:solution_gas_over_time_acoustic} also give the `algebraic growth rate' of the backward reaction,
\begin{equation}
    \label{eq:algebraic_growth_rate_gas_acoustic}
    \gamma_{2} = \frac{\dtg}{2 \tau} \frac{\cs \, \hrd}{\tug (t = 0)}.
\end{equation}

\subsubsection{It grows because the drag force has just the right pattern to accelerate the flows of a sound wave.}
\label{ssub:1D_backward_block_3}

To understand why the sound wave grows, we need to remember that the gas acts as a harmonic oscillator (Eq.~\ref{eq:forcing_in_kernel_of_operator_gas_acoustic}). It is well known that when a harmonic oscillator is forced at its natural frequency, the system selects the mode that has the same phase as the forcing to grow. Following that insight, we draw in Fig.~\ref{fig:Impact_of_dust_feedback_acoustic} the state of the system when the gas pattern has zero phase-shift with respect to the dust pattern, ${ \varphi = 0}$. We see that the perturbed drag force is everywhere in the same direction as the perturbed velocities of the right-propagating sound wave. Therefore, the gas accelerates and the wave gets stronger.

A resonant sound wave moves with the dust advective mode, meaning it can maintain ${ \varphi = 0 }$ at all times. Therefore, at any given instant ${ t }$, it is getting strengthened. Consequently, it grows algebraically.

Non-resonant waves see their phase-shift evolve in time. At a given instant ${ t }$ they may have a phase-shift of zero. If they have the right perturbed velocity pattern, then we are in the situation of Fig.~\ref{fig:Impact_of_dust_feedback_acoustic}, and the waves gain amplitude. But a little while later, they will have ${ \varphi = \pi }$ and lose amplitude. These two situations happen equally often, explaining why non-resonant waves do not grow. 

This applies, in particular, to left-propagating sound waves. Indeed, they have the same velocity perturbation pattern as the right-propagating waves. Because of this symmetry, we intuit that the perturbed drag force tends to share its effort equally between the two modes. We conjecture that this explains the factor ${ 1 / \, 2 }$ in Eqs.~\eqref{eq:algebraic_growth_rate_gas_acoustic}.

\subsection{Combining the two processes: exponential instability}
\label{sub:1D_combine}

The two preceding subsections described the two halves of the 1D acoustic \RDI. In \S\ref{sub:1D_forward} we saw that a sound wave can concentrate dust, and in \S\ref{sub:1D_backward} we saw that a dust density perturbation can drive a specific sound wave. Intuitively, if the favoured wave turns out to be the sound wave that created the density perturbation in the first place, we can expect a feedback loop between the two mechanisms, leading to a full-scale exponential instability. The present subsection offers to formalise this intuition. 

Since we expect an exponential instability, let us fully Fourier-decompose each variable, ${ f(t, x) = \widehat{f} \, \e^{i (k x - \omega t)} }$. The governing equations are those of Eqs.~\eqref{eq:perturbation_equations_acoustic}, and the dispersion relation is given by Eq.~\eqref{eq:dispersion_relation_acoustic_1D}.  As we saw in \S\ref{sub:dispersion_relation}, the dispersion relation is enough to estimate the growth rate~\eqref{eq:growth_rate_from_dispersion_relation_acoustic} of the acoustic \RDI, but less helpful in understanding its physics. To make progress on that front, let us look closely at the eigenvectors. Singular perturbation theory \citep[e.g.][]{Seyranian03} suggests a solution of the form
\begin{alignat}{5}
    & \omega      &&= \omega_{0}      &&+ \sqrt{\dtg} \, \omega_{1}     &&+ \dtg \, \omega_{2} &&+ ... \, , \label{eq:Puiseux_expansion_acoustic} \\
    & \widehat{f} &&= \widehat{f}_{0} &&+ \sqrt{\dtg} \widehat{f}_{1}   &&+ \dtg \, \widehat{f}_{2} &&+ ... \, . \nonumber
\end{alignat}
where ${ \widehat{f} }$ represents any of ${ \hrg, \hug, \hrd, \hud }$, and where ${ \widehat{\varrho}_{g, 0} }$, ${ \widehat{u}_{g, 0} }$ and ${ \widehat{u}_{d, 0} }$ are all equal to zero, but not ${ \omegaO }$ nor ${ \widehat{\varrho}_{d, 0}}$. The latter assumption means that our mode, from the outset, possesses a comparatively large dust density.

We will also assume that the dimensionless wave-number ${ K = \cs k \tau }$ is of order unity, so that the sizes of all the terms in the perturbation equations~\eqref{eq:perturbation_equations_acoustic} are dictated by the dust-to-gas ratio ${ \dtg }$. We shall relax this assumption in \S\ref{sec:small_and_large_scale}. 

\subsubsection{Dust equations at order ${ 1 }$}

At order ${ 1 }$, the dust equations (\ref{eq:perturbation_equations_acoustic_continuity_dust}, \ref{eq:perturbation_equations_acoustic_momentum_dust}) only give 
\begin{equation}
    \label{eq:order_0_dust_acoustic}
    \omegaO = v k ,
\end{equation}
which shows that, to leading order, the mode frequency equals the dust advective frequency. This means our ansatz (in which the perturbation in dust density dominates) isolates the dust advective mode of \S\ref{sub:dispersion_relation}. Our perturbative approach helps us study how this mode is modified when we leave the test-particle limit.

\subsubsection{Gas equations at order ${ \sqrt{\dtg} }$}

At this order, the gas equations (\ref{eq:perturbation_equations_acoustic_continuity_gas}, \ref{eq:perturbation_equations_acoustic_momentum_gas}) give, after some basic algebra\footnote{Recall that the zeroth order gas terms are null.}
\begin{subequations}
    \label{eq:order_half_gas_acoustic}
    \begin{align}
        & v \, \hrgO = \pm \cs \, \hrgO , \label{eq:order_half_gas_acoustic_1} \\
        & \hugO = \cs \, \hrgO . \label{eq:order_half_gas_acoustic_2}
    \end{align}
\end{subequations}
The first equation confirms that our ansatz in powers of ${ \sqrt{\dtg} }$ holds only at resonance, ${ v = \cs }$. Out of resonance, the only non-trivial terms are in powers of ${ \dtg }$, as per standard perturbation theory. The second equation tells us that, to leading order, the gas perturbation takes the form of a sound wave.

\subsubsection{Dust equations at order ${ \sqrt{\dtg} }$}
\label{ssub:1D_combine_block_3}

It is at this order that things get interesting, because we start to recover the instability mechanisms discussed in \S\ref{sub:1D_forward} and \S\ref{sub:1D_backward}. Indeed, the dust equations (\ref{eq:perturbation_equations_acoustic_continuity_dust}, \ref{eq:perturbation_equations_acoustic_momentum_dust}) give
\begin{subequations}
    \label{eq:order_half_dust_acoustic}
    \begin{align}
        -i \omegaI \, \hrdO &= -i \, k \hudO , \label{eq:order_half_dust_acoustic_continuity} \\
        0 &= - \frac{1}{\tau} (\hudO - \hugO) . \label{eq:order_half_dust_acoustic_momentum}
    \end{align}
\end{subequations}
These equations are identical to Eqs.~\eqref{eq:IVP_dust_acoustic_v2}, once one has waited for a few stopping times so that the dust has reached its terminal velocity. From there we deduce
\begin{equation}
    \label{eq:order_half_dust_acoustic_v2}
    -i \omegaI \, \hrdO = -i \, k \hugO ,
\end{equation}
which is analogous to Eq.~\eqref{eq:algebraic_growth_rate_dust_acoustic}. This equation captures the dust-accumulation mechanism of \S\ref{sub:1D_forward} and Fig.~\ref{fig:Dust_accumulation_acoustic}, which relies on the divergence of the sound wave's flow. 

When discussed earlier, this forward process was isolated; here it is connected to the backward reaction, and that connection is made clear through the neat asymptotic ordering, as we see in the next paragraph.

\subsubsection{Gas equations at order ${ \dtg }$}

At order ${ \dtg }$, the gas equations (\ref{eq:perturbation_equations_acoustic_continuity_gas}, \ref{eq:perturbation_equations_acoustic_momentum_gas}) give
\begin{subequations}
    \label{eq:order_one_gas_acoustic}
    \begin{align}
        &  -i \omegaI \, \hrgO - i \, \cs k \, \hrgI = -i \, k \hugI , \label{eq:order_one_gas_acoustic_continuity} \\
        &  -i \omegaI \, \hugO - i \, \cs k \, \hugI = -i \cs^{2} k \, \hrgI + \frac{1}{\tau} v \, \hrdO , \label{eq:order_one_gas_acoustic_momentum}
    \end{align}
\end{subequations}
which are similar to Eqs.~\eqref{eq:IVP_gas_acoustic}, thereby justifying our assumption from \S\ref{sub:1D_backward} that the dominant drag contribution comes from the dust's density perturbation. One can eliminate the second-order gas terms, leading to
\begin{subequations}
    \label{eq:order_one_gas_acoustic_v2}
    \begin{align}
        &  -i \omegaI \, \hrgO = \frac{1}{2 \tau} \, \hrdO , \label{eq:order_one_gas_acoustic_v2_density} \\
        &  -i \omegaI \, \hugO = \frac{\cs}{2 \tau} \, \hrdO. \label{eq:order_one_gas_acoustic_v2_velocity}
    \end{align}
\end{subequations}
These equations reveal that the gas perturbations are forced by the dust-density perturbation, exactly as described in \S\ref{sub:1D_backward} and Fig.~\ref{fig:Impact_of_dust_feedback_acoustic}. 
Once again, this growth process was isolated in \S\ref{sub:1D_backward}, but is now connected to the forward action.

\subsubsection{Bringing it all together}
\label{ssub:1D_combine_block_5}

Combining Eq.~\eqref{eq:order_half_dust_acoustic_v2} with Eq.~\eqref{eq:order_one_gas_acoustic_v2_velocity} gives us
\begin{align}
    \sqrt{\dtg} \, \omega_{1, \pm} =& \pm \sqrt{ - \frac{-i \, k \hugO}{\hrdO} \times \frac{\dtg}{2 \tau} \frac{\cs \, \hrdO}{\hugO} } = \pm \sqrt{- \gamma_{1} \gamma_{2}} \nonumber \\
    {} =& \pm \sqrt{ \frac{ i \dtg \cs k}{2 \tau} } , \label{eq:growth_rate_acoustic_RDI}
\end{align}
Note that ${ \gamma_{1} }$ and ${ \gamma_{2} }$ appear because ${ \hrdO }$ and ${ \hugO }$ are the initial amplitudes of the dust density perturbation and of the gas velocity perturbation, respectively.

Since ${ \omega_{1, +} }$ has a positive real part, this equation shows a full instability that grows exponentially with time. This is of course the 1D acoustic \RDI, and we recognise the eigenfrequency from Eq.~\eqref{eq:growth_rate_from_dispersion_relation_acoustic} of this paper, Eq.~13 of \citetalias{SquireHopkins18b} or Eqs.~15 and 19 of \citetalias{HopkinsSquire18a}. Again, it is only valid when the dimensionless wave-number ${ K = \cs k \tau }$ is of order unity, and at resonance.

What is new and interesting is the un-factorised form of the equation, which shows that the exponential growth rate of the \RDI\ is the geometric mean of the `algebraic growth rates' of its constituent instabilities from \S\ref{sub:1D_forward} and \S\ref{sub:1D_backward}. This structure shows nicely that the mechanism behind the 1D acoustic \RDI\ is a feedback loop between the forward action and the backward reaction. A sound wave concentrates dust, and the ensuing dust density perturbation strengthens the sound wave, which can then concentrate dust faster, etc... Note that the idea of a feedback loop was already hinted at by \citetalias{SquireHopkins20} in the context of the \SI, and \citetalias{HopkinsSquire18a} in the context of the acoustic \RDI. Our main contribution here is to rigorously formalise the idea.

Recall that the forward mechanism is most efficient when the dust pattern is phase-shifted by ${ - \pi / 2 }$ with respect to the sound wave, whereas the backward mechanism is most efficient when there is no phase-shift. Equation \eqref{eq:order_one_gas_acoustic_v2_density} gives
\begin{equation}
    \label{eq:combine_acoustic}
    \hrdO = \sqrt{2 \cs k \tau} \, \e^{ - i \pi / 4} \, \hrgO ,
\end{equation}
which means that the dust pattern is phase-shifted by ${ - \pi / 4 }$ with respect to the sound wave. This is a middle ground where both mechanisms can operate sufficiently well for the feedback loop to work.

The un-factorised growth rate also elucidates why growth is faster on small scales. Sound waves concentrate dust by moving it from the pressure peak to the green oval of Fig.~\ref{fig:Dust_accumulation_acoustic}. The dust speed is determined by the gas speed as per Eq.~\eqref{eq:solution_dust_velocity_over_time_acoustic}. Consequently, sound waves of similar amplitude impart the same speed to the dust. However, shorter-wavelength sound waves have less distance to cover between the pressure peak and the green oval, so they concentrate dust faster.

\section{Multi-dimensional case}
\label{sec:2D}

Extending the previous discussion to 3D takes surprisingly little effort. The key is to realise that even in 3D, all that is relevant happens within the same single dimension, i.e. along the direction of sound wave propagation. \\

First, we decompose the solutions to Eqs.~\eqref{eq:perturbation_equations_acoustic} into spatial Fourier modes of the form ${ f(t, \bx) = \widehat{f} \, \e^{i (\bk \cdot \bx - \omega t)} }$. Then, we decompose the drift vector $\bv$ into its components parallel and perpendicular to the wave-vector ${ \bk }$,
\begin{equation} 
    \label{eq:decompose_v}
    \bv = \vpara \mathbf{\hat{k}} + \bvperp , \,\,\,\,\,\,\, \text{ where } \,\,\,\,\,\,\, \bvperp \bcdot \bk = 0 .
\end{equation}
We can perform the same decomposition for the perturbed gas and dust velocities. The perturbation equations \eqref{eq:perturbation_equations_acoustic} become
\begin{subequations}
    \label{eq:perturbation_equations_acoustic_3D_para}
    \begin{align}
        & \pD t \trg = - i \, k \tugpara , \label{eq:perturbation_equations_acoustic_3D_para_continuity_gas} \\
        & \pD t \tugpara =  - i \cs^{2} k \, \trg + \frac{\dtg}{\tau} (\tudpara - \tugpara) + \frac{\dtg \vpara}{\tau} (\trd - \trg) , \label{eq:perturbation_equations_acoustic_3D_para_momentum_gas} \\
        & \pD t \trd + i \, \vpara k \, \trd = - i \, k \tudpara , \label{eq:perturbation_equations_acoustic_3D_continuity_dust} \\
        & \pD t \tudpara + i \, \vpara k \, \tudpara =  - \frac{1}{\tau} (\tudpara - \tugpara) , \label{eq:perturbation_equations_acoustic_3D_para_momentum_dust}
    \end{align}
\end{subequations}
in the parallel direction and
\begin{subequations}
    \label{eq:perturbation_equations_acoustic_3D_perp}
    \begin{align}
        & \pD t \tbugperp = \frac{\dtg}{\tau} (\tbudperp - \tbugperp) + \frac{\dtg}{\tau}  \bvperp (\trd - \trg) , \label{eq:perturbation_equations_acoustic_3D_perp_momentum_gas} \\
        & \pD t \tbudperp + i \, \vpara k \, \tbudperp = - \frac{1}{\tau} (\tbudperp - \tbugperp) . \label{eq:perturbation_equations_acoustic_3D_perp_momentum_dust}
    \end{align}
\end{subequations}
in the perpendicular directions. The interesting thing is that the parallel perturbed velocities together with the relative perturbations in densities are governed by the closed system ~\eqref{eq:perturbation_equations_acoustic_3D_para}, which is exactly the system governing the 1D case. Therefore, everything we said about the 1D acoustic \RDI\ applies to the 3D acoustic \RDI. Simply, the resonance condition becomes ${ \vpara = \cs }$, and ${ \bvperp }$ is unconstrained.

The perpendicular drift does not impact the instability mechanism because it can only move a dust particle within its waveplane (cf.\ Section 2.5). As such, in the forward action, ${ \mathbf{v_{\perp}} }$ is unable to transport the dust away from the velocity node, and in the backward reaction it cannot change the drag forces impacting the sound wave. Yet, ${ \mathbf{v_{\perp}} }$ being unconstrained is not without consequence. It makes the acoustic \RDI\ much more viable. In 1D it needs ${ v = \cs }$, but in 3D it only needs ${ v \geq \cs }$ to work.

To verify that nothing untoward happens in the perpendicular direction, consider the complete dispersion relation of the 3D system. It can be written as 
\begin{multline}
    \label{eq:dispersion_relation_acoustic_3D}
    \left\{ (\omega^{2} - \cs^{2} k^{2}) (\omega - \vpara k) (\omega - \vpara k + \frac{i}{\tau}) + \dtg \, P(\omega, \vpara k, \cs k, \frac{1}{\tau}) \right\} \\
    \times \left\{ \omega \, (\omega - \vpara k + \frac{i}{\tau}) + \dtg \, \frac{i \, (\omega - \vpara k)}{\tau} \right\}^{2} = 0 .
\end{multline}
The first factor, which represents the modes in densities and parallel velocities, recovers the 1D dispersion relation \eqref{eq:dispersion_relation_acoustic_1D} and therefore the 1D \RDI. The second factor, which represents the modes in perpendicular velocities, manifests two damped modes that are hence of no interest to us. This confirms that all the unstable dynamics are covered by the 1D case.

\section{Discussion and conclusion}
\label{sec:conclusion}

This paper is the first of a pair aimed at understanding how \RDIs\ develop. They are a class of instabilities that are likely to emerge in a range of astrophysical settings, from \AGNs\ to molecular clouds. In proto-planetary discs, in particular, an \RDI\ called the \SI\ is thought to play an important role by bridging the metre-gap of planet formation. Unfortunately, because of the dynamical complexity of two coupled fluids (gas and dust), \RDIs\ are still poorly understood. This confusion makes it hard to evaluate the robustness of these instabilities, to figure out how they saturate, or to interpret simulations. 

In the present paper, we focus on the acoustic \RDI. We chose to consider the simplest \RDI\ first for pedagogical reasons, as going slowly through an elementary example often helps form an intuition for the general case.

We find that the acoustic \RDI\ is built on two constituent mechanisms. If they could be separated from each other, these two processes would both become algebraically unstable at resonance. The fact they are connected leads to an exponentially unstable feedback loop. This interpretation was first presented in \citetalias{SquireHopkins20} for the `epicyclic \RDI'. First, a sound wave is able to concentrate dust because it is composed of convergent flows and because drag forces the dust to follow these flows (\S\ref{ssub:1D_forward_block_2}). Concurrently, the nascent dust accumulation exerts a back-reaction onto the gas; its excess drag force on the gas has the right pattern to strengthen the flows of a sound wave, and thereby the wave itself (\S\ref{ssub:1D_backward_block_2}). The stronger wave can now concentrate dust faster, leading to stronger clumps, which make the wave even stronger, and so on...

We find that resonance is helpful to prevent the dust drifting away from the velocity node it has been brought to by the gas drag (\S\ref{ssub:1D_forward_block_3}), and to make sure the sound wave is always strengthened by the excess drag from the dust clump, and never weakened (\S\ref{ssub:1D_backward_block_3}). But we stress that perfect resonance is only necessary for the algebraic instabilities of the constituent mechanisms; it is inessential for the exponential \RDI, which is robust to detuning (\S\ref{sec:detuning}).

We formalise the feedback loop idea with a clear asymptotic theory. Subsection~\ref{sub:1D_combine} offers a new way to compute the growth rate of any particular \RDI. Our procedure is equivalent to the linear algebra of Squire \& Hopkins, but it trades efficiency for transparency. This makes it easier to understand the physics behind a given \RDI.

Our focus has been on using the acoustic \RDI\ as a simple example on which to build an intuition for how other \RDIs\ develop. But the acoustic \RDI\ is of great interest on its own and could emerge in many astrophysical objects. Research in that area has only just started but looks promising. In the dusty torus of an \AGN, the radiation pressure is so strong that the dust's relative drift becomes supersonic. This makes the acoustic \RDI\ possible, which might explain the clumps, velocity sub-structures, and turbulence observed in these regions \citepalias{HopkinsSquire18a}. In the \ISM, a cluster of massive stars can cause enough radiation pressure to accelerate the surrounding dust to supersonic speeds relative to the gas, thus allowing the acoustic \RDI\ to work. This might source turbulence in the surrounding clouds, and create fluctuations in the dust-to-gas ratio -- which could in turn modify their chemistry and thermodynamics, and possibly explain the observed variability in stellar composition \citepalias{HopkinsSquire18a}. Furthermore, once self-gravity is taken into account, the acoustic \RDI\ may facilitate the formation of molecular clouds, or create small (sub-Jeans size) dust-rich areas inside clouds \citep{Zhuravlev21}. In the winds of cool stars, it is unclear whether the radiation pressure is powerful enough to make the dust supersonic. There might be some regions of the wind where it is, and some where it is not. In the regions where the acoustic \RDI\ is active, it would develop into turbulence and might explain observed clumpy sub-structures. In the regions where it is inactive, large-scale structures such as arcs and shells could subsist \citepalias{HopkinsSquire18a}.

There are many physical effects that might be important in the aforementioned astrophysical contexts but that are neglected in the present paper. We have already mentioned self-gravity, and we cover a few more in the appendices. In \S\ref{sec:detuning} we study the instability off-resonance, and find that it maintains a high growth rate even for moderate detunings. In \S\ref{sec:variable_stopping_times} we study how the instability develops when the grains' stopping time varies with the local gas density or velocity. Nevertheless, our analysis remains limited to an idealised environment where all the dust grains have the same size. \cite{Squire+22} use simulations to relax that hypothesis, and find that the smallest grains in a distribution do not exhibit \RDI-like behaviour. The present work also does not include any magnetic field or background turbulence, which might severely limit the applicability of its results to molecular clouds \citep{PriestleyWhitworth22}. Finally, our approach is by design limited to the linear regime. \cite{Moseley+19} study the non-linear regime of the acoustic \RDI\ with numerical experiments, and find that it saturates in a turbulent state whose dust is concentrated in filaments, jets, plumes or fronts -- depending on the dust's stopping time and the dust-to-gas ratio.

A follow-up paper will explain how the more important but more complicated \SI, which is the \RDI\ arising from inertial waves, grows. Understanding this instability is important, and has already been the focus of considerable effort (\citealt{YoudinJohansen07, Jacquet+11, LinYoudin17}, \citetalias{SquireHopkins20}).

\newpage
\section*{Acknowledgments}

We wish to thank our anonymous referee as well as V. V. Zhuravlev for their comments that helped us clarify the manuscript. Support for N.M. was provided by a Cambridge International \& Isaac Newton Studentship.

\section*{Data Availability}

No new data were generated or analysed in support of this research.

\bibliographystyle{mnras}
\bibliography{main}

\begin{thebibliography}{}
\makeatletter
\relax
\def\mn@urlcharsother{\let\do\@makeother \do\$\do\&\do\#\do\^\do\_\do\%\do\~}
\def\mn@doi{\begingroup\mn@urlcharsother \@ifnextchar [ {\mn@doi@}
  {\mn@doi@[]}}
\def\mn@doi@[#1]#2{\def\@tempa{#1}\ifx\@tempa\@empty \href
  {http://dx.doi.org/#2} {doi:#2}\else \href {http://dx.doi.org/#2} {#1}\fi
  \endgroup}
\def\mn@eprint#1#2{\mn@eprint@#1:#2::\@nil}
\def\mn@eprint@arXiv#1{\href {http://arxiv.org/abs/#1} {{\tt arXiv:#1}}}
\def\mn@eprint@dblp#1{\href {http://dblp.uni-trier.de/rec/bibtex/#1.xml}
  {dblp:#1}}
\def\mn@eprint@#1:#2:#3:#4\@nil{\def\@tempa {#1}\def\@tempb {#2}\def\@tempc
  {#3}\ifx \@tempc \@empty \let \@tempc \@tempb \let \@tempb \@tempa \fi \ifx
  \@tempb \@empty \def\@tempb {arXiv}\fi \@ifundefined
  {mn@eprint@\@tempb}{\@tempb:\@tempc}{\expandafter \expandafter \csname
  mn@eprint@\@tempb\endcsname \expandafter{\@tempc}}}

\bibitem[\protect\citeauthoryear{{Andrews}}{{Andrews}}{2020}]{Andrews20}
{Andrews} S.~M.,  2020, \mn@doi [Annual Review of Astronomy and Astrophysics]
  {10.1146/annurev-astro-031220-010302}, 58, 483

\bibitem[\protect\citeauthoryear{{Chan}, {Manger}, {Li}, {Yang}, {Zhu},
  {Armitage}  \& {Ho}}{{Chan} et~al.}{2022}]{Chan+22}
{Chan} Y.-M.,  {Manger} N.,  {Li} Y.,  {Yang} C.-C.,  {Zhu} Z.,  {Armitage}
  P.~J.,   {Ho} S.,  2022, \mn@doi [arXiv e-prints]
  {10.48550/arXiv.2210.02339}, \href
  {https://ui.adsabs.harvard.edu/abs/2022arXiv221002339C} {p. arXiv:2210.02339}

\bibitem[\protect\citeauthoryear{{Chiang} \& {Youdin}}{{Chiang} \&
  {Youdin}}{2010}]{ChiangYoudin10}
{Chiang} E.,  {Youdin} A.~N.,  2010, \mn@doi [Annual Review of Earth and
  Planetary Sciences] {10.1146/annurev-earth-040809-152513}, \href
  {https://ui.adsabs.harvard.edu/abs/2010AREPS..38..493C} {38, 493}

\bibitem[\protect\citeauthoryear{{Decin}}{{Decin}}{2021}]{Decin21}
{Decin} L.,  2021, \mn@doi [Annual Review of Astronomy and Astrophysics]
  {10.1146/annurev-astro-090120-033712}, 59, 337

\bibitem[\protect\citeauthoryear{{Deguchi}}{{Deguchi}}{1997}]{Deguchi97}
{Deguchi} S.,  1997, in Planetary Nebulae. Proceedings of the 180th Symposium
  of the International Astronomical Union (IAU).
p.~151

\bibitem[\protect\citeauthoryear{{Hickox} \& {Alexander}}{{Hickox} \&
  {Alexander}}{2018}]{HickoxAlexander18}
{Hickox} R.~C.,  {Alexander} D.~M.,  2018, \mn@doi [Annual Review of Astronomy
  and Astrophysics] {10.1146/annurev-astro-081817-051803}, 56, 625

\bibitem[\protect\citeauthoryear{{Hopkins} \& {Squire}}{{Hopkins} \&
  {Squire}}{2018}]{HopkinsSquire18a}
{Hopkins} P.~F.,  {Squire} J.,  2018, \mn@doi [\mnras] {10.1093/mnras/sty1982},
  \href {https://ui.adsabs.harvard.edu/abs/2018MNRAS.480.2813H} {480, 2813}

\bibitem[\protect\citeauthoryear{{Hughes}, {Duch\^{e}ne}  \&
  {Matthews}}{{Hughes} et~al.}{2018}]{Hughes+18}
{Hughes} A.~M.,  {Duch\^{e}ne} G.,   {Matthews} B.~C.,  2018, \mn@doi [Annual
  Review of Astronomy and Astrophysics] {10.1146/annurev-astro-081817-052035},
  56, 541

\bibitem[\protect\citeauthoryear{{Jacquet}, {Balbus}  \& {Latter}}{{Jacquet}
  et~al.}{2011}]{Jacquet+11}
{Jacquet} E.,  {Balbus} S.,   {Latter} H.,  2011, \mn@doi [\mnras]
  {10.1111/j.1365-2966.2011.18971.x}, \href
  {https://ui.adsabs.harvard.edu/abs/2011MNRAS.415.3591J} {415, 3591}

\bibitem[\protect\citeauthoryear{{Johansen} \& {Youdin}}{{Johansen} \&
  {Youdin}}{2007}]{JohansenYoudin07}
{Johansen} A.,  {Youdin} A.,  2007, \mn@doi [\apj] {10.1086/516730}, \href
  {https://ui.adsabs.harvard.edu/abs/2007ApJ...662..627J} {662, 627}

\bibitem[\protect\citeauthoryear{{Johansen}, {Youdin}  \& {Mac Low}}{{Johansen}
  et~al.}{2009}]{Johansen+09}
{Johansen} A.,  {Youdin} A.,   {Mac Low} M.-M.,  2009, \mn@doi [\apjl]
  {10.1088/0004-637X/704/2/L75}, \href
  {https://ui.adsabs.harvard.edu/abs/2009ApJ...704L..75J} {704, L75}

\bibitem[\protect\citeauthoryear{{Lambrechts}, {Johansen}, {Capelo}, {Blum}  \&
  {Bodenschatz}}{{Lambrechts} et~al.}{2016}]{Lambrechts+16}
{Lambrechts} M.,  {Johansen} A.,  {Capelo} H.~L.,  {Blum} J.,   {Bodenschatz}
  E.,  2016, \mn@doi [\aap] {10.1051/0004-6361/201526272}, \href
  {https://ui.adsabs.harvard.edu/abs/2016A&A...591A.133L} {591, A133}

\bibitem[\protect\citeauthoryear{{Lin}}{{Lin}}{2021}]{Lin21}
{Lin} M.-K.,  2021, \mn@doi [\apj] {10.3847/1538-4357/abcd9b}, \href
  {https://ui.adsabs.harvard.edu/abs/2021ApJ...907...64L} {907, 64}

\bibitem[\protect\citeauthoryear{{Lin} \& {Youdin}}{{Lin} \&
  {Youdin}}{2017}]{LinYoudin17}
{Lin} M.-K.,  {Youdin} A.~N.,  2017, \mn@doi [\apj] {10.3847/1538-4357/aa92cd},
  \href {https://ui.adsabs.harvard.edu/abs/2017ApJ...849..129L} {849, 129}

\bibitem[\protect\citeauthoryear{Luongo}{Luongo}{1995}]{Luongo95}
Luongo A.,  1995, \mn@doi [Journal of Sound and Vibration]
  {https://doi.org/10.1006/jsvi.1995.0387}, 185, 377

\bibitem[\protect\citeauthoryear{{Morris}}{{Morris}}{1993}]{Morris93}
{Morris} M.,  1993, in European Southern Observatory Conference and Workshop
  Proceedings. p.~60

\bibitem[\protect\citeauthoryear{{Moseley}, {Squire}  \& {Hopkins}}{{Moseley}
  et~al.}{2019}]{Moseley+19}
{Moseley} E.~R.,  {Squire} J.,   {Hopkins} P.~F.,  2019, \mn@doi [\mnras]
  {10.1093/mnras/stz2128}, \href
  {https://ui.adsabs.harvard.edu/abs/2019MNRAS.489..325M} {489, 325}

\bibitem[\protect\citeauthoryear{{Priestley} \& {Whitworth}}{{Priestley} \&
  {Whitworth}}{2022}]{PriestleyWhitworth22}
{Priestley} F.~D.,  {Whitworth} A.~P.,  2022, \mn@doi [\mnras]
  {10.1093/mnras/stac627}, \href
  {https://ui.adsabs.harvard.edu/abs/2022MNRAS.512.1407P} {512, 1407}

\bibitem[\protect\citeauthoryear{{Seyranian} \& {Mailybaev}}{{Seyranian} \&
  {Mailybaev}}{2003}]{Seyranian03}
{Seyranian} A.~P.,  {Mailybaev} A.~A.,  2003, Multiparameter stability theory
  with mechanical applications.
 Series on stability, vibration, and control of systems, Series A Vol. 13,
  World Scientific, Singapore ; River Edge, NJ

\bibitem[\protect\citeauthoryear{{Squire} \& {Hopkins}}{{Squire} \&
  {Hopkins}}{2018a}]{SquireHopkins18a}
{Squire} J.,  {Hopkins} P.~F.,  2018a, \mn@doi [\mnras] {10.1093/mnras/sty854},
  \href {https://ui.adsabs.harvard.edu/abs/2018MNRAS.477.5011S} {477, 5011}

\bibitem[\protect\citeauthoryear{{Squire} \& {Hopkins}}{{Squire} \&
  {Hopkins}}{2018b}]{SquireHopkins18b}
{Squire} J.,  {Hopkins} P.~F.,  2018b, \mn@doi [\apjl]
  {10.3847/2041-8213/aab54d}, \href
  {https://ui.adsabs.harvard.edu/abs/2018ApJ...856L..15S} {856, L15}

\bibitem[\protect\citeauthoryear{{Squire} \& {Hopkins}}{{Squire} \&
  {Hopkins}}{2020}]{SquireHopkins20}
{Squire} J.,  {Hopkins} P.~F.,  2020, \mn@doi [\mnras]
  {10.1093/mnras/staa2311}, \href
  {https://ui.adsabs.harvard.edu/abs/2020MNRAS.498.1239S} {498, 1239}

\bibitem[\protect\citeauthoryear{{Squire}, {Moroianu}  \& {Hopkins}}{{Squire}
  et~al.}{2022}]{Squire+22}
{Squire} J.,  {Moroianu} S.,   {Hopkins} P.~F.,  2022, \mn@doi [\mnras]
  {10.1093/mnras/stab3377}, \href
  {https://ui.adsabs.harvard.edu/abs/2022MNRAS.510..110S} {510, 110}

\bibitem[\protect\citeauthoryear{{Toschi} \& {Bodenschatz}}{{Toschi} \&
  {Bodenschatz}}{2009}]{ToschiBodenschatz09}
{Toschi} F.,  {Bodenschatz} E.,  2009, \mn@doi [Annual Review of Fluid
  Mechanics] {10.1146/annurev.fluid.010908.165210}, \href
  {https://ui.adsabs.harvard.edu/abs/2009AnRFM..41..375T} {41, 375}

\bibitem[\protect\citeauthoryear{Weingartner \& Draine}{Weingartner \&
  Draine}{2001}]{Weingartner01}
Weingartner J.~C.,  Draine B.~T.,  2001, \mn@doi [The Astrophysical Journal]
  {10.1086/320963}, 553, 581

\bibitem[\protect\citeauthoryear{{Youdin} \& {Goodman}}{{Youdin} \&
  {Goodman}}{2005}]{YoudinGoodman05}
{Youdin} A.~N.,  {Goodman} J.,  2005, \mn@doi [\apj] {10.1086/426895}, \href
  {https://ui.adsabs.harvard.edu/abs/2005ApJ...620..459Y} {620, 459}

\bibitem[\protect\citeauthoryear{{Youdin} \& {Johansen}}{{Youdin} \&
  {Johansen}}{2007}]{YoudinJohansen07}
{Youdin} A.,  {Johansen} A.,  2007, \mn@doi [\apj] {10.1086/516729}, \href
  {https://ui.adsabs.harvard.edu/abs/2007ApJ...662..613Y} {662, 613}

\bibitem[\protect\citeauthoryear{{Zhuravlev}}{{Zhuravlev}}{2019}]{Zhuravlev19}
{Zhuravlev} V.~V.,  2019, \mn@doi [\mnras] {10.1093/mnras/stz2390}, \href
  {https://ui.adsabs.harvard.edu/abs/2019MNRAS.489.3850Z} {489, 3850}

\bibitem[\protect\citeauthoryear{{Zhuravlev}}{{Zhuravlev}}{2021}]{Zhuravlev21}
{Zhuravlev} V.~V.,  2021, \mn@doi [\mnras] {10.1093/mnras/staa3424}, \href
  {https://ui.adsabs.harvard.edu/abs/2021MNRAS.500.2209Z} {500, 2209}

\bibitem[\protect\citeauthoryear{{van Dishoeck}}{{van
  Dishoeck}}{2004}]{vanDishoeck04}
{van Dishoeck} E.~F.,  2004, \mn@doi [Annual Review of Astronomy and
  Astrophysics] {10.1146/annurev.astro.42.053102.134010}, 42, 119

\makeatother
\end{thebibliography}

\appendix

\newpage
\section{The acoustic RDI off-resonance}
\label{sec:detuning}

In an infinite domain, acoustic waves of all wave-vectors can propagate, so the resonance condition ${ \bk \bcdot \bv = \cs k }$ will always be satisfied by one of them (provided ${ v \geq \cs }$). But in a bounded domain, there is a geometric restriction on which modes can exist, and it will often prohibit the sound wave that has just the right wave-vector for perfect resonance. This problem is certain to impact numerical simulations, unless the dimensions of the simulation domain are chosen perfectly. Hence, it is crucial to study the case of imperfect tuning between the drift and the wave.

To do so, we introduce the dimensionless parameter ${ \delta }$ so that ${ \bv \bcdot \bk = [1 + \delta] \, \cs k }$ i.e. ${\vpara = [1 + \delta] \, \cs }$ or (in 1D) ${M = 1 + \delta}$ (recall that $M$ is the Mach number). This single scalar parameter captures the key physics of \S\ref{sec:2D}: strictly speaking, detuning is a misdirection of the wave-vector from perfect resonance, but because the instability is fundamentally one-dimensional, for all intents and purposes we can think of detuning as a mismatch of the sound speed.

With this convenient notation, the 1D Fourier-transformed perturbation equations become
\begin{subequations}
    \label{eq:1D_Fourier_transformed_perturbation_equations_acoustic_with_detuning}
    \begin{align}
        & - i \omega \, \hrg \! = \! - i \, k \hug , \label{eq:1D_Fourier_transformed_perturbation_equations_acoustic_with_detuning_continuity_gas} \\
        & - i \omega \, \hug \! = \! - i \cs^{2} k \, \hrg \! + \! \frac{\dtg}{\tau} (\hud \! - \! \hug) \! + \! \frac{\dtg \cs}{\tau} [1 \! + \! \delta] (\hrd - \hrg) , \label{eq:1D_Fourier_transformed_perturbation_equations_acoustic_with_detuning_momentum_gas} \\
        & - i \omega \, \hrd \,+ i \, [1 + \delta] \cs k \, \hrd = - i \, k \hud , \label{eq:1D_Fourier_transformed_perturbation_equations_acoustic_with_detuning_continuity_dust} \\
        & - i \omega \, \hud + i \, [1 + \delta] \cs k \, \hud =  - \frac{1}{\tau} (\hud - \hug) . \label{eq:1D_Fourier_transformed_perturbation_equations_acoustic_with_detuning_momentum_dust}
    \end{align}
\end{subequations}
Notice that these are the equations at perfect resonance, plus three extra terms due to detuning. This type of `perturbed nearly defective' system of equations has been studied by \cite{Luongo95}. In their terminology, ${ \dtg }$ would be the `perturbation parameter' and ${ \delta }$ the `mistuning parameter'. But we prefer the word `detuning'.

\subsection{Mathematical approach}
\label{sub:detuning_approach}

Since we are interested in the regime where both parameters ($\mu$, $\delta$) are small, we could expand any variable ${ f }$ into
\begin{equation}
    \label{eq:Expansion_detuning_acoustic_two_parameters}
    f = f_{0, 0} + \sqrt{\dtg} f_{0, 1} + \delta f_{1, 0} + \dtg f_{0, 2} + \delta \sqrt{\dtg} f_{1, 1} + \delta^{2} f_{0, 2} + ...
\end{equation}
But \cite{Luongo95} advises to select a curve that starts from the origin of parameter space, ${ \{\dtg, \delta\} = \{0, 0\} }$, and to reduce the 2-variable perturbation problem to a 1-variable perturbation problem along this curve. They find that this approach, which is a form of distinguished limit, has better convergence properties than the naive double-limit approach.

We introduce the family of curves ${ \delta = \xi \sqrt{\dtg} }$, where the slope parameter ${ \xi }$ can take any real value. Note that $\xi$ does not measure distance along a given curve, but rather labels the different curves. This way, every point of parameter space is described by one of our curves. On a given curve, we can expand ${ f }$ into
\begin{equation}
    f = f_{0} (\xi) + \sqrt{\dtg} f_{1} (\xi) + \dtg f_{2} (\xi) + ... \label{eq:Expansion_detuning_after_curve_selection}
\end{equation}
The next step is to distinguish three regimes of detuning: small detuning when ${ |\xi| \ll 1 }$, intermediate detuning when ${ |\xi| \sim 1 }$ and large detuning when ${ |\xi| \gg 1 }$.

\begin{figure*}
    \centering
    \includegraphics[width = \textwidth]{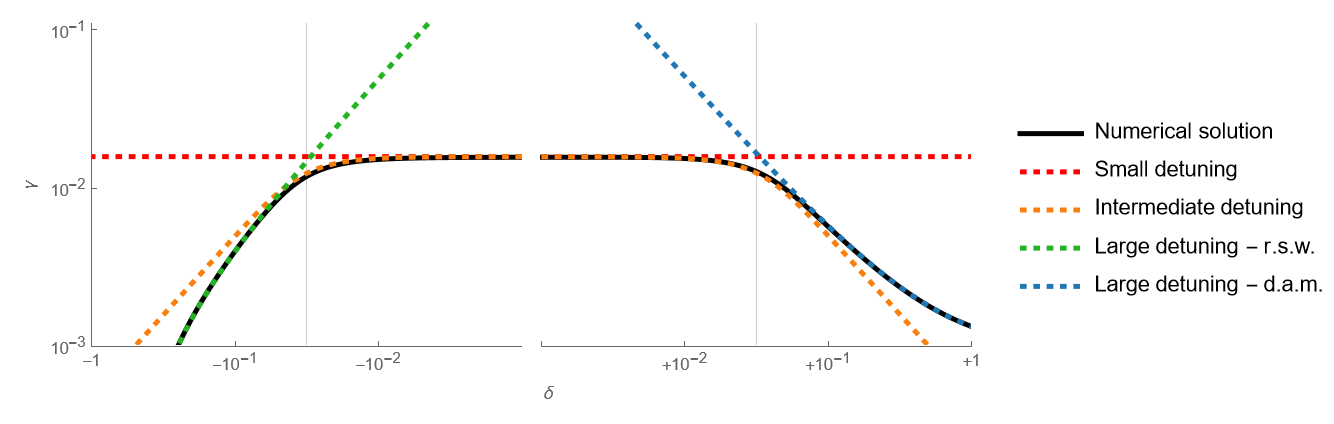}
    \caption{Growth rate of the instability ${ \gamma }$ as a function of the detuning parameter ${ \delta }$. The solid black line displays the exact growth rate, as deduced from Eq.~\eqref{eq:1D_Fourier_transformed_perturbation_equations_acoustic_with_detuning}. The dashed red line represents the prediction from Eq.~\eqref{eq:growth_rate_acoustic_RDI}. It should be valid at low detuning, when ${ |\delta| \ll \dtg^{1/2} }$. This limit is shown by the thin grey vertical lines. The dashed orange line corresponds to the prediction from Eq.~\eqref{eq:growth_rate_acoustic_RDI_with_detuning}, which should be valid at intermediate detuning, when ${ |\delta| \sim \dtg^{1/2} }$. The dashed blue and green lines are the predictions from standard perturbation theory, as given by Eqs.~\eqref{eq:growth_rates_from_standard_perturbation_theory_acoustic_with_detuning}. They should be accurate at large detuning, when ${ |\delta| \gg \dtg^{1/2} }$.
    \textit{Parameters}: ${ \cs = 1 }$, ${ \tau = 1 }$, ${ k = 1 }$, ${ \dtg = 10^{-3} }$.
    }
    \label{fig:Effect_of_detuning}
\end{figure*}

\subsection{Small detuning}
\label{sub:small_detuning}

If we follow \S\ref{sub:1D_combine} and use the ansatz in powers of ${ \sqrt{\dtg} }$ from Eq.~\eqref{eq:Puiseux_expansion_acoustic}, we find that in the regime of small detuning, ${ |\xi| \ll 1 }$, the new term in the gas momentum equation~\eqref{eq:1D_Fourier_transformed_perturbation_equations_acoustic_with_detuning_momentum_gas} is of size ${ \mu \, \delta \ll \dtg^{3/2} }$. But recall that the hierarchical procedure of \S\ref{sub:1D_combine} reveals that, in the gas equations, only terms of lower order than ${ \dtg }$ are relevant to the acoustic \RDI. As such, the new term in the gas momentum equation is negligible at small detuning. Concurrently, the new term in the dust continuity equation is of size ${ \delta \ll \sqrt{\dtg} }$ and the one in the dust momentum equation is of size ${ \sqrt{\mu} \delta \ll \dtg }$, but \S\ref{sub:1D_combine} tells us that in the dust equations, only the terms of lower order than ${ \sqrt{\dtg} }$ are significant.

In summary, in the regime of small detuning, all the new terms are negligible at all relevant orders. As a consequence, the governing equations are still Eqs.~(\ref{eq:order_0_dust_acoustic}, \ref{eq:order_half_gas_acoustic}, \ref{eq:order_half_dust_acoustic}, \ref{eq:order_one_gas_acoustic}), and the growth rate of Eq.~\eqref{eq:growth_rate_acoustic_RDI} remains valid. This finding connects well to the discussion on resonance angle in \S 3.7.1.2 of \citetalias{HopkinsSquire18a}.

We think it is an important result that the growth rate keeps its maximum value over a fairly large part of wave-vector space (see Fig.~\ref{fig:Effect_of_detuning}). It makes simulations of the instability much easier to set up than one could have feared. Having a large enough box and a reasonable resolution should be enough to ensure that the sampling of wave-vector space contains at least one point within the large unstable region. There should be no need to fine-tune the box's dimensions.

\subsection{Intermediate detuning}
\label{sub:intermediate_detuning}

At intermediate detuning, ${ |\xi| \sim 1 }$, the new term in the dust continuity equation~\eqref{eq:1D_Fourier_transformed_perturbation_equations_acoustic_with_detuning_continuity_dust} attains a magnitude of ${ \sqrt{\dtg} }$. As such, it is big enough to play a role in the hierarchy of \S\ref{sub:1D_combine}. It significantly impacts  the forward action and thus the discussion of \S\ref{ssub:1D_combine_block_3}. On the mathematical side of things, it modifies Eqs.~\eqref{eq:order_half_dust_acoustic} so that
\begin{subequations}
    \label{eq:order_half_dust_acoustic_with_detuning}
    \begin{align}
        -i \omegaI \, \hrdO + i \, \xi \, \cs k \, \hrdO &= -i \, k \hudO , \label{eq:order_half_dust_acoustic_with_detuning_continuity} \\
        0 &= \frac{1}{\tau} (\hudO - \hugO) , \label{eq:order_half_dust_acoustic_with_detuning_momentum}
    \end{align}
\end{subequations}
from which one quickly obtains
\begin{equation}
    \label{eq:growth_rate_forward_action_acoustic_with_detuning}
        \left( - i \omegaI + i \, \xi \, \cs k \right) \, \hrdO = - i \, k \hugO ,
\end{equation}
which replaces Eq.~\eqref{eq:order_half_dust_acoustic_v2}. When combined with Eq.~\eqref{eq:order_one_gas_acoustic_v2_velocity}, we obtain
\begin{equation}
    \label{eq:growth_rate_acoustic_RDI_with_detuning}
        \omega_{1, \pm} = \pm \sqrt{\frac{i \cs k}{2 \tau} + \left( \frac{\delta \cs k}{2 \sqrt{\mu}} \right)^{2} } + \frac{\delta, \cs k}{2 \sqrt{\mu}} ,
\end{equation}
where we have replaced ${ \xi }$ by ${ \delta / \sqrt{\mu} }$ so that the final expression is written in terms of physical parameters. The imaginary part of ${ \sqrt{\dtg} \, \omega_{1, +} }$ gives an approximation for the growth rate of the full instability. It is valid at intermediate detuning by design, but also at small detuning. Essentially, the growth rate becomes smaller with detuning because some dust that has been brought to the green oval of Fig.~\ref{fig:Dust_accumulation_acoustic} by the sound wave's flows manages to drift away from the oval.

\subsection{Large detuning}
\label{sub:large_detuning}

The regime of large detuning, ${ |\xi| \gg 1 }$, is easier to study because the frequencies of the sound wave and of the dust advective mode are sufficiently different that the mode collision highlighted in \S\ref{sub:dispersion_relation} becomes irrelevant. As such, the growth rates of the instability can be estimated via standard perturbation theory. 

Let us write the perturbation equations~\eqref{eq:1D_Fourier_transformed_perturbation_equations_acoustic_with_detuning} in the matrix form 
\begin{equation}
    \label{eq:perturbation_equations_matrix_form}
    \left\{ \mathbf{M_{0}} + \dtg \mathbf{M_{1}} \right\} \mathbf{\zeta} = i \omega \mathbf{\zeta} ,
\end{equation}
where ${ \mathbf{M_{0}} }$ and ${ \mathbf{M_{1}} }$ are matrices and ${ \mathbf{\zeta} }$ is the solution vector. Note that all of these (and ${ \omega }$) are functions of ${ \xi }$. Standard perturbation theory states that if we consider a mode that is characterised in the $\mu=0$ regime by the frequency ${ \omegaO }$, the right mode structure vector ${ \mathbf{\zeta_{0}^{\text{R.}}} }$ and the left mode structure vector ${ \mathbf{\zeta_{0}^{\text{L.}}} }$ (such that ${ \mathbf{M_{0}} \mathbf{\zeta_{0}^{\text{R.}}} = i \omegaO \mathbf{\zeta_{0}^{\text{R.}}} }$ and ${ \mathbf{\zeta_{0}^{\text{L.}}} \mathbf{M_{0}} = i \omegaO \mathbf{\zeta_{0}^{\text{L.}}} }$), then when $\mu\neq 0$ the mode's frequency becomes ${ \omega = \omegaO + \dtg \omega_{1} + ... }$, with
\begin{equation}
    \label{eq:standard_perturbation_theory}
     \omega_{1} = - i \frac{\mathbf{\zeta_{0}^{\text{L.}}} \mathbf{M_{1}} \mathbf{\zeta_{0}^{\text{R.}}}}{\mathbf{\zeta_{0}^{\text{L.}}} \mathbf{\zeta_{0}^{\text{R.}}}} .
\end{equation} 
The imaginary part of this perturbed frequency is the growth rate of the instability.

We know from \S\ref{sub:dispersion_relation} that in the regime of test particles there are two sound waves, a dust advective mode, and a damped dust mode. The acoustic \RDI\ only involves the right-propagating sound wave and the dust advective mode. When we apply Eq.~\eqref{eq:standard_perturbation_theory} to these, we find
\begin{subequations}
    \label{eq:growth_rates_from_standard_perturbation_theory_acoustic_with_detuning}
    \begin{align}
        & \mathfrak{Im} (\omega_{1}^{\text{rsw}})  = \frac{- 1}{2 \tau} \left\{ 2 + \delta + \frac{1}{\delta} \frac{1}{1 + \left[ \delta \cs k \tau \right]^{2}} \right\} , \label{eq:growth_rates_from_standard_perturbation_theory_acoustic_with_detuning_right_sound_wave} \\
        & \mathfrak{Im} (\omega_{1}^{\text{dam}})  = \frac{1}{\tau \delta} \frac{(1 + \delta)^{2}}{2 + \delta} . \label{eq:growth_rates_from_standard_perturbation_theory_acoustic_with_detuning_dust_advective_mode}
    \end{align}
\end{subequations}
where the superscript ${ ^{\text{rsw}} }$ denotes the right-propagating sound wave and ${ ^{\text{dam}} }$ the dust advective mode.

Note that Eq.~\eqref{eq:growth_rates_from_standard_perturbation_theory_acoustic_with_detuning_dust_advective_mode} corresponds to Eq.~13 of \citetalias{HopkinsSquire18a}, who call it the ``non-resonant quasi-drift mode". Eq.~\eqref{eq:growth_rates_from_standard_perturbation_theory_acoustic_with_detuning_right_sound_wave} partially matches Eq.~10 and the ``quasi-sound mode" of \citetalias{HopkinsSquire18a}, though the correspondence is complicated by the fact that we use an expansion in $\dtg$ whereas they use an expansion in $K^{-1}$. Certainly, our estimate performs better when ${ \dtg^{1/2} \ll |\delta| \ll 1 }$. \\

Figure~\ref{fig:Effect_of_detuning} shows in blue and green the estimate of the growth rate provided by formulas~\eqref{eq:growth_rates_from_standard_perturbation_theory_acoustic_with_detuning}. It is good everywhere except near the region of mode collision, where one needs to use the estimate from Eq.~\eqref{eq:growth_rate_acoustic_RDI}, shown in red. As expected, the formula~\eqref{eq:growth_rate_acoustic_RDI_with_detuning} provides a better prediction (shown in orange) at intermediate detuning than either resonant perturbation theory or standard perturbation theory.

\section{The acoustic RDI at small and large scales}
\label{sec:small_and_large_scale}

In the main body of the paper, we assumed that the dimensionless wave-number $K$ is of order unity. Let us relax this assumption and study how the instability changes when the wavelength is large, or small.

To simplify this auxiliary analysis, we shall only consider eigenvalues not eigenvectors. This allows us to work straight from the dispersion relation~\eqref{eq:dispersion_relation_acoustic_1D}. Its dimensionless form is
\begin{equation}
    \label{eq:dispersion_relation_acoustic_1D_adimensional}
    (\Omega - K)^{2} (\Omega + K) (\Omega - K + i) + \dtg \, P(\Omega, K) = 0,
\end{equation}
with ${ \Omega = \omega \tau }$ the dimensionless frequency and
\begin{equation}
    \label{eq:P_acoustic_adimensional}
    P(\Omega, K) = i \left[ \Omega^{3} - K \Omega^{2} - K^{2} \Omega + K^{3} - i K^{2} \right] \!.
\end{equation}

\begin{figure}
    \centering
    \includegraphics[width = 1.0 \linewidth]{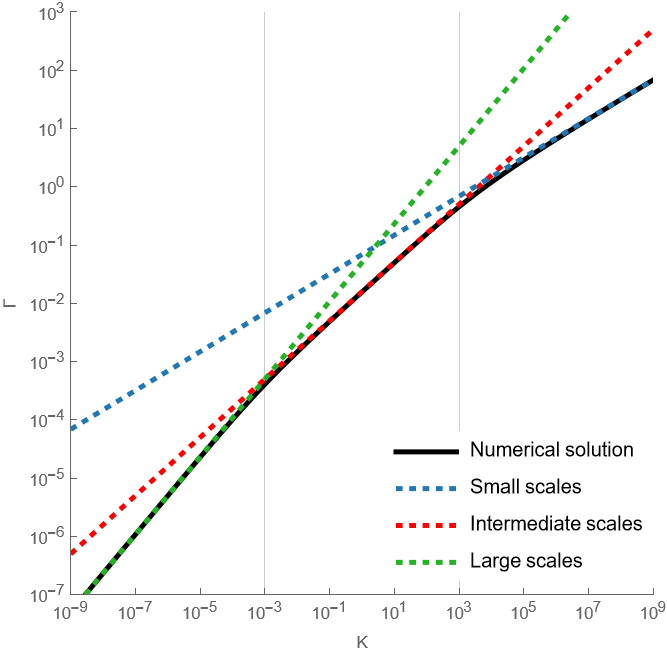}
    \caption{
        The dimensionless growth rate of the instability ${ \Gamma = \gamma \tau }$ as a function of the dimensionless wave-number ${ K = \cs k \tau }$. The solid black line displays the exact growth rate, as deduced from Eq.~\eqref{eq:dispersion_relation_acoustic_1D}. 
        The dashed red line represents the prediction on intermediate scales from Eq.~\eqref{eq:growth_rate_from_dispersion_relation_acoustic_v2}. 
        The dashed blue line is the prediction om small scales, i.e. Eq.~\eqref{eq:growth_rate_acoustic_1D_adimensional_small_scale}.
        The dashed green line is the prediction on long scales, i.e. Eq.~\eqref{eq:growth_rate_acoustic_1D_adimensional_large_scale}. 
        The thin vertical grey lines indicate the boundaries between regimes, i.e. $K= \mu,\,1/\mu$. 
        \textit{Parameters}: ${ M = 1 }$, ${ \dtg = 10^{-3} }$.}
    \label{fig:Effect_of_K}
\end{figure}

Guided by the numerical solution (see Fig.~\ref{fig:Effect_of_K}), we distinguish three regimes: large scales (${K \ll \dtg}$), intermediate scales (${\dtg \ll K \ll \dtg^{-1}}$) and small scales (${\dtg^{-1} \ll K}$). The intermediate regime is the focus of \S\ref{sub:dispersion_relation}, we already know that the instability grows at the rate given by Eq.~\eqref{eq:growth_rate_from_dispersion_relation_acoustic_v2} at those scales. Let us consider the extreme regimes now.

\subsection{Large scales}
\label{ssub:large_scales}

In this regime, $K$ is very small. This allows us to neglect most terms in the dispersion relation. To do so rigorously, let us expand the frequency in powers of ${\dtg^{1/3}}$ as
\begin{equation}
    \label{eq:decomposition_frequency_large_scales}
    \Omega = K + \mu^{1/3} \, \Omegap + \dtg^{2/3} ... ,
\end{equation}
where ${ \Omegap }$ is a complex whose modulus is of order unity. This scaling with $\dtg^{1/3}$ is the only valid one in the regime where ${ K \ll \dtg \ll 1 }$. At order $\dtg^{1}$, the dispersion relation~\eqref{eq:dispersion_relation_acoustic_1D_adimensional} simplifies to
\begin{equation}
    \label{eq:dispersion_relation_acoustic_1D_adimensional_large_scale_develloped_simplified}
    \Omegap^{3} = i \, K^{2},
\end{equation}
which has three solutions. Two of these have a positive imaginary part, leading to the dimensionless growth rate
\begin{equation}
    \label{eq:growth_rate_acoustic_1D_adimensional_large_scale}
    \Gamma = \gamma \tau = \frac{1}{2} K^{2/3} \dtg^{1/3} + \mathcal{O} (\dtg^{2/3}).
\end{equation}
This connects to Eq.~9 of \citetalias{HopkinsSquire18a}, who call it the ``long-wavelength pressure-free mode".

By tracing back the origin of the l.h.s. term in Eq.~\eqref{eq:dispersion_relation_acoustic_1D_adimensional_large_scale_develloped_simplified}, we learn that this instability is a triadic interaction between the two sound waves and the dust advective mode. This is in agreement with the discussion in \S5.5 of \cite{Zhuravlev21}, without being in direct disagreement with the claim in \S3.1.2 of \citetalias{HopkinsSquire18a} that this instability is non-resonant. Indeed, our numerical inquiry showed a slightly higher growth rate for $M \neq 1$.

\subsection{Small scales}
\label{ssub:small_scales}

In this regime, the numerical solution to Eq.~\eqref{eq:dispersion_relation_acoustic_1D_adimensional} indicates that $\Gamma$ and $\Omega$ diverge (see Fig.~\ref{fig:Effect_of_K}). We regularise things by dividing the dispersion relation by $K^{4}$. This gives an equation for ${ \tOmega = \Omega / K }$:
\begin{equation}
\label{eq:dispersion_relation_acoustic_1D_adimensional_small_scales_regularised}
    (\tOmega - 1)^{2} (\tOmega + 1) (\tOmega - 1 + i K^{-1}) + \dtg \, P(\tOmega, K^{-1}) = 0,
\end{equation}
where
\begin{equation}
    \label{eq:P_acoustic_adimensional_small_scales_regularised}
    P(\tOmega, K^{-1}) = i K^{-1} \left[ \tOmega^{3} - \tOmega^{2} - \tOmega + 1 - i K^{-1} \right]\!.
\end{equation}
From here, we can follow much the same steps as in \S\ref{ssub:large_scales}. We expand the regularised dimensionless frequency as
\begin{equation}
    \label{eq:decomposition_frequency_small_scales}
    \tOmega = 1 + \dtg^{1/3} \, \tOmegap + \dtg^{2/3} ...,
\end{equation}
which is the only valid scaling when ${ K^{-1} \ll \dtg \ll 1 }$. At order ${\dtg^{1}}$, the dispersion relation~\eqref{eq:dispersion_relation_acoustic_1D_adimensional_small_scales_regularised} simplifies to
\begin{equation}
    \label{eq:dispersion_relation_acoustic_1D_adimensional_small_scales_develloped_simplified}
    \tOmega^{'3} = - \frac{1}{2} K^{-2},
\end{equation}
which has three solutions. One of them has a positive imaginary part, leading to an instability of regularised dimensionless growth rate
\begin{equation}
\label{eq:growth_rate_acoustic_1D_adimensional_small_scale_regularised}
    \tilde{\Gamma} = \sqrt{3} \left( \frac{K^{-2} \dtg}{16} \right)^{1/3} + \mathcal{O} (\dtg^{2/3})
\end{equation}
and actual dimensionless growth rate
\begin{equation}
    \label{eq:growth_rate_acoustic_1D_adimensional_small_scale}
    \Gamma = \sqrt{3} \left( \frac{K \dtg}{16} \right)^{1/3} + \mathcal{O} (\dtg^{2/3}) .
\end{equation}
This connects to Eq.~14 of \citetalias{SquireHopkins18b} and Eq.~16 of \citetalias{HopkinsSquire18a}, who call it the ``short-wavelength resonant quasi-drift mode".

Once again, by tracing the origin of the l.h.s. term in Eq.~\eqref{eq:dispersion_relation_acoustic_1D_adimensional_small_scales_develloped_simplified}, we learn that the small-scale instability is a triadic interaction. But this time, it involves the right-propagating sound wave, the dust advective mode and the damped dust mode. 

Figure~\ref{fig:Effect_of_K} shows how well the growth rates provided by Eqs.~\eqref{eq:growth_rate_acoustic_1D_adimensional_large_scale}, \eqref{eq:growth_rate_from_dispersion_relation_acoustic_v2} and \eqref{eq:growth_rate_acoustic_1D_adimensional_small_scale} compare to the numerical solution of Eq.~\eqref{eq:dispersion_relation_acoustic_1D}. It is an extension of Fig.~1 in \citetalias{SquireHopkins18b}.

\section{Dependence of the stopping time on density and velocity }
\label{sec:variable_stopping_times}

So far, we have used an arbitrary and constant stopping time ${ \tau }$ for the dust. This assumption is valid in some regimes. For instance, a large particle at low Reynolds is in the Stokes regime of drag, where the stopping time is \citep{ChiangYoudin10}
\begin{equation}
    \label{eq:stopping_time_Stokes}
    \tau \approx \frac{4 \rho_{p} R_{p}^{2} \sigma_{g}}{9 \cs}  .
\end{equation}
Here ${ \rho_{p} }$ is the density of the typical dust particle, ${ R_{p} }$ is its radius, and ${ \sigma_{g} }$ is the collision cross-section of the typical gas particle. We can reasonably assume these quantities do not vary. The stopping time is thus only a function of the sound speed ${ \cs }$, which is constant for our isothermal gas.

But in some other regimes, the assumption of constant stopping time is invalid. For instance, dust particles that are smaller than the gas's mean free path are in the Epstein regime of drag. At the high Mach numbers required by the acoustic \RDI, \citetalias{HopkinsSquire18a} introduce the modified Epstein drag law
\begin{equation}
    \label{eq:stopping_time_Epstein_high_Mach}
    \tau \approx \frac{\rho_{p} R_{p}}{\rho_{g} \cs} \left( 1 + a \frac{|\bud - \bug|^{2}}{\cs^{2}} \right)^{- 1/2} ,
\end{equation}
where ${ a \approx 0.22 }$ for an isothermal gas. Notice how the stopping time is now dependent on the gas density, ${ \rho_{g} }$, and on the relative velocity between dust and gas, ${ |\bud - \bug| }$. 

In this appendix we explore  how these variations in the stopping time impact the acoustic \RDI\ and its physical picture. It shows how easy investigations into an \RDI's robustness become thanks to the procedure of \S\ref{sub:1D_combine}.

\subsection{Mathematical approach}
\label{sub:variable_stopping_time_mathematical_approach}

Following the notation of \citetalias{SquireHopkins18b}, we account for the perturbation in stopping time by
\begin{equation}
    \label{eq:variations_in_stopping_time}
    \frac{\tau'}{\overline{\tau}} = - \zeta_{\rho} \, \rgi - \zeta_{v} \, \frac{(\budi - \bugi) \bcdot \bv}{v^{2}} ,
\end{equation}
where ${ \zeta_{\rho} = 1 }$ and ${ \zeta_{v} = a \, v^{2} / \left(\cs^{2} + a \, v^{2}\right) }$ are scalar parameters, hereafter called `sensitivities'. Note that they would take different values if the gas was not isothermal.

They bring two new terms in each of the linearised momentum equations~(\ref{eq:perturbation_equations_acoustic_momentum_gas}, \ref{eq:perturbation_equations_acoustic_momentum_dust}):
\begin{subequations}
    \label{eq:perturbation_equations_acoustic_with_varying_stopping_time}
    \begin{align}
        & \pD t \bugi =  
        \begin{multlined}[t]
            - \cs^{2} \, \grad \rgi + \frac{\dtg}{\tau} \left[ \mathbf{\mathcal{I}} \! + \! \zeta_{v} \frac{\bv}{v} \! \otimes \! \frac{\bv}{v} \right] (\budi \! - \! \bugi) \\
            + \frac{\dtg}{\tau} \bv \, (\rdi - \left[ 1 - \zeta_{\rho} \right] \rgi) ,
        \end{multlined}
        \label{eq:perturbation_equations_acoustic_with_varying_stopping_time_momentum_gas} \\
        & \pD t \budi \! + \! \bv \! \bcdot \! \grad \budi \! = \! - \frac{1}{\tau} \left[ \mathbf{\mathcal{I}} \! + \! \zeta_{v} \frac{\bv}{v} \! \otimes \! \frac{\bv}{v} \right] (\budi \! - \! \bugi) \! - \! \frac{\zeta_{\rho}}{\tau} \bv \rgi , \label{eq:perturbation_equations_acoustic_with_varying_stopping_time_momentum_dust}
    \end{align}
\end{subequations}
where ${ \mathbf{\mathcal{I}} }$ is the identity matrix and ${ \otimes }$ the outer product.

The ${ \zeta_{v} }$ term brings a noteworthy difficulty. It breaks the neat decoupling of \S\ref{sec:2D} by making the parallel problem dependent on the perpendicular velocities.

\subsection{One-dimensional case}
\label{sub:variable_stopping_time_1D}

Let us first consider the 1D system. Since it has no perpendicular direction, we can forget about the aforementioned coupling for now.

When we follow the hierarchy of \S\ref{sub:1D_combine}, we immediately see that all modifications to the gas momentum equation~\eqref{eq:perturbation_equations_acoustic_with_varying_stopping_time_momentum_gas} only involve terms that are too small to play a significant role in the instability. The modifications to the dust momentum equation~\eqref{eq:perturbation_equations_acoustic_with_varying_stopping_time_momentum_dust} do, however, impact the forward action. We find that Eqs.~\eqref{eq:order_half_dust_acoustic} are replaced by 
\begin{subequations}
    \label{eq:order_half_dust_acoustic_with_varying_stopping_time_1D}
    \begin{align}
        -i \omegaI \, \hrdO &= -i \, k \hudO , \label{eq:order_half_dust_acoustic_with_varying_stopping_time_1D_continuity} \\
        0 &= - \frac{1 + \zeta_{v}}{\tau} (\hudO - \hugO) - \frac{\zeta_{\rho}}{\tau} \cs \, \hrgO . \label{eq:order_half_dust_acoustic_with_varying_stopping_time_1D_momentum}
    \end{align}
\end{subequations}

Let us first consider the term brought by the dependence of the stopping time on gas density. To study its influence in isolation, let us imagine that ${ \zeta_{v} = 0 }$. Since the structure of a sound wave is such that ${ \hrgO = \hugO / \cs }$, the two gaseous terms in Eq.~\eqref{eq:order_half_dust_acoustic_with_varying_stopping_time_1D_momentum} cancel each other perfectly. This means that the dust decouples from the gas. 

Figure~\ref{fig:Dust_accumulation_acoustic} helps understand why. The structure of a right-propagating sound wave is such that the gas velocity bump is also a gas density bump. At this position, the gas flows are directed to the right and drag the dust in the same direction. But the gas over-density locally increases the drag coefficient, so the dust slows down its drift in the lab frame, or equivalently drifts to the left in the frame of Fig.~\ref{fig:Dust_accumulation_acoustic}. The two effects cancel each other, and the dust does not move -- and \textit{a fortiori} does not clump.

This would kill the forward action and the \RDI, but ${ \zeta_{v} }$ is not null. So let us analyze the term brought by the dependence of the stopping time on dust-to-gas relative velocity. This term strengthens the incentive for the dust to follow the gas flows. As such, the perfect balance we described in the previous paragraph, between the gas flows dragging the dust to the right and the over-density helping the dust drift to the left, is broken. ${ \zeta_{v} }$ gives the flows the upper hand, and dust accumulates in the green oval once more, although the `algebraic growth rate' of the forward action is reduced to 
\begin{equation}
    \label{eq:growth_rate_forward_action_with_varying_stopping_time_1D}
    \gamma_{1} = \left( 1 - \frac{\zeta_{\rho}}{1 + \zeta_{v}} \right) \frac{- i \, k \hug}{\hrd} .
\end{equation}

Using the formula~\eqref{eq:growth_rate_acoustic_RDI} which states that the exponential growth rate of the \RDI\ is the geometric mean of the algebraic growth rates of its two constituent mechanisms, we obtain
\begin{equation}
    \label{eq:growth_rate_acoustic_RDI_with_varying_stopping_time_1D}
    \omega_{1, \pm} = \pm \sqrt{1 - \frac{\zeta_{\rho}}{1 + \zeta_{v}}} \sqrt{ \frac{i \cs k}{2 \tau} } .
\end{equation}
This result is confirmed by an asymptotic analysis of the dispersion relation (\textit{cf.} \S\ref{sub:dispersion_relation}), and is in agreement with Eq.~13 of \citetalias{SquireHopkins18b}.

\subsection{Multi-dimensional case}
\label{sub:variable_stopping_time_2D}

In the more realistic multi-dimensional case, we have to deal with the coupling of parallel and perpendicular motions.

Its first effect is that it modifies the relative sizes of the different variables' Fourier amplitudes. In the Stokes regime, the ansatz in powers of ${ \sqrt{\dtg} }$ from Eq.~\eqref{eq:Puiseux_expansion_acoustic} started at order ${ \dtg^{0} }$ for ${ \omega }$ and ${ \hrd }$; at order ${ \dtg^{1/2} }$ for ${ \hrg }$, ${ \hug }$ and ${ \hud }$; and at order ${ \dtg^{1} }$ for ${ \hbugperp }$ and ${ \hbudperp }$. In the Epstein regime, the ansatz must start at order ${ \dtg^{1/2} }$ for ${ \hbudperp }$.

Its second effect is that, at each step of the hierarchy of \S\ref{sub:1D_combine}, we need to consider 3 equations (density, parallel velocity \& perpendicular velocity) instead of 2. Fortunately, according to the ansatz, the perturbed perpendicular motions are slow. We find that the new equations (and the new terms in the usual equations) play no role at order 1 in the dust, order ${ \sqrt{\dtg} }$ in the gas, or order ${ \dtg }$ in the gas. They only play a role at order ${ \sqrt{\dtg} }$ in the dust. 

This means that the variations in stopping time only impact the forward action. We find that Eqs.~\eqref{eq:order_half_dust_acoustic} are replaced by
\begin{subequations}
    \label{eq:order_half_dust_acoustic_with_varying_stopping_time_3D}
    \begin{align}
        & -i \omegaI \, \hrdO = -i \, k \hudparaO , \label{eq:order_half_dust_acoustic_with_varying_stopping_time_3D_continuity} \\
        & 0 = 
        \begin{multlined}[t]
            - \frac{\left[1 + \zeta_{v} \left(\vpara / v\right)^{2}\right]}{\tau} (\hudparaO - \hugparaO) - \frac{\zeta_{\rho}}{\tau} \vpara \, \hrgO \\ 
            - \frac{\zeta_{v}}{\tau} \frac{\vpara}{v} \frac{\bvperp}{v} \bcdot \hbudperpO ,
        \end{multlined}
        \label{eq:order_half_dust_acoustic_with_varying_stopping_time_3D_momentum_para} \\
        & \mathbf{0} =
        \begin{multlined}[t]
            - \frac{1}{\tau} \left[\mathbf{\mathcal{I}} + \zeta_{v} \frac{\bvperp}{v} \otimes \frac{\bvperp}{v}\right] \hbudperpO - \frac{\zeta_{\rho}}{\tau} \bvperp \, \hrgO \\ 
            - \frac{\zeta_{v}}{\tau} \frac{\vpara}{v} \frac{\bvperp}{v} (\hudparaO -     \hugparaO) .
        \end{multlined}
        \label{eq:order_half_dust_acoustic_with_varying_stopping_time_3D_momentum_perp}
    \end{align}
\end{subequations}

Admittedly, this looks complicated. But one can use Eq.~\eqref{eq:order_half_dust_acoustic_with_varying_stopping_time_3D_momentum_perp} to eliminate ${ \bvperp \bcdot \hbudperpO }$. This reduces the system to
\begin{subequations}
    \label{eq:order_half_dust_acoustic_with_varying_stopping_time_3D_v2}
    \begin{align}
        -i \omegaI \, \hrdO &= -i \, k \hudparaO , \label{eq:order_half_dust_acoustic_with_varying_stopping_time_3D_v2_continuity} \\
        0 &= - \frac{1 + \chi_{v}}{\tau} (\hudparaO - \hugparaO) - \frac{\chi_{\rho}}{\tau} \cs \, \hrgO , \label{eq:order_half_dust_acoustic_with_varying_stopping_time_3D_v2_momentum_para}
    \end{align}
\end{subequations}
which is analogous to the set of equations~\eqref{eq:order_half_dust_acoustic_with_varying_stopping_time_1D} governing the 1D forward action with varying stopping time. Simply, we replace the true sensitivities ${ \left\{ \zeta_{\rho}, \, \zeta_{v} \right\} }$ with the effective sensitivities
\begin{equation}
    \label{eq:effective_sensitivities_to_variations_in_stopping_time}
    \chi_{\rho} = \frac{\zeta_{\rho}}{1 + (\vperp / v)^{2}} , \,\,\,\,\,\,\,\,\,\,\,\, \chi_{v} = \frac{\zeta_{v} \, (\vpara / v)^{2}}{1 + (\vperp / v)^{2}} .
\end{equation}

Therefore, the growth rate of the multi-dimensional acoustic \RDI\ in the high-Mach Epstein regime is
\begin{equation}
    \label{eq:growth_rate_acoustic_RDI_with_varying_stopping_time_2D}
    \omega_{1, \pm} = \pm \sqrt{1 \! - \! \frac{\chi_{\rho}}{1 + \chi_{v}}} \sqrt{ \frac{i \cs k}{2 \tau} } = \pm \sqrt{1 \! - \! \frac{\zeta_{\rho}}{1 + \zeta_{v}}} \sqrt{ \frac{i \cs k}{2 \tau} } ,
\end{equation}
where the last equality is obtained after a few lines of basic algebra. This formula is in perfect agreement with Eq.~13 of \citetalias{SquireHopkins18b}, and Eq.~15 of \citetalias{HopkinsSquire18a}.

\bsp	
\label{lastpage}
\end{document}